\documentclass{revtex4-1}

\usepackage{graphicx}
\usepackage{float}
\usepackage{amsmath,amssymb}

\begin{document}
\title{ Detailed Analysis of  p+p Elastic Scattering Data in the Quark-Diquark Model of Bialas and Bzdak \\ from  $\sqrt{s}=23.5$ GeV to 7 TeV }

\author{F. Nemes}
\affiliation{Institute of Physics, E\"otv\"os University \\ P\'azm\'any P. s. 1/A, H-1117 Budapest, Hungary}
\email{frigyes.janos.nemes@cern.ch}

\author{T. Cs\"org\H{o}}
\affiliation{Wigner Research Centre for Physics, RMKI \\ H-1525 Budapest 114, P.O.Box 49, Hungary}
\email{csorgo.tamas@wigner.mta.hu}

\begin{abstract}
Final results of a detailed analysis of p+p elastic scattering data
are presented,
 utilizing the quark-diquark model of protons in a form  proposed by Bialas and Bzdak.
The differential cross-section of elastic proton-proton collisions 
is analyzed in a detailed and systematic manner at small momentum transfers,
starting from the energy range of CERN ISR at  $ \sqrt{s}= 23.5 $ GeV,
including also recent TOTEM data at the present LHC energies at $\sqrt{s} = 7$ TeV. 
These studies confirm the picture that the size of proton increases systematically 
with increasing energies, while the size of the constituent quarks and diquarks
remains approximately independent of (or only increases only slightly with) the colliding energy. 
The detailed analysis indicates correlations between model parameters and also indicates 
an increasing role of shadowing at LHC energies.
Within the investigated class of models, a simple and model-independent phenomenological relation was discovered that connects the total p+p 
scattering cross-section to the effective quark, diquark size and their average separation. Our best fits  indicate, that the relative error 
of this phenomenological relation is 10-15 \% in the considered energy range.
\end{abstract}

\maketitle
\newpage

\section{Introduction}
The differential cross section of elastic scattering of p+p collisions allows one to 
study the internal structure of protons using the theory of diffraction. 
Varying the momentum transfer one can change the resolution of the investigation: 
increasing the momentum transfer corresponds to looking more and more deeply inside the
structure of protons. One of the fundamental outcomes of diffractive p+p scattering
studies was the indication that protons have a finite size and a complicated
internal structure, thus the protons can be considered as composite objects.

Our interest in this problem has been triggered by two factors:
an interesting series of recent theoretical work and also new data from
the TOTEM experiment at CERN LHC. These are detailed below.\par 
Recently, we became aware of an 
inspiring series of papers of Bialas, Bzdak and collaborators, who studied
elastic proton-proton~\cite{Bialas:2006qf}, 
pion-proton~\cite{Bzdak:2007qq}
and nucleus-nucleus collisions~\cite{Bialas:2006kw,Bialas:2007eg}
in a framework where the proton was considered as a composite object that contains
correlated quark and diquark constituents. 
In this work, we confirm their main conclusion:
the quark-diquark model of nucleon structure at low momentum transfer does 
capture the main features of this problem and indeed it deserves a closer, more detailed attention.
The first results that form the basis of our current study were
made public in a proceeding material~\cite{Nemes:2012mq}. 
In that study we already reported
about  several details of the present investigation.  In contrast to the paper of
Bialas and Bzdak in ref.~\cite{Bialas:2006qf}, we 
not only included an estimation of the best values of the model parameters,
but also determined their errors and also the fit quality. \par
In this paper, we refined our first results ~\cite{Nemes:2012mq} 
by decreasing the number of fit parameters from five to three which leads to a far
better error estimation, and less correlations among the  remaining three model parameters.
In order to reach these goals,
we utilized standard experimental techniques, such as multi-parameter 
optimalization or fitting with the help of the  MINUIT function minimalization
and multi-parameter optimalization package~\cite{James:1975dr}.

In addition to these simple, straightforward and
interesting theoretical investigations of elastic scattering data from 
CERN ISR in the energy range of 
$\sqrt{s} =$ 23.5, 30.7, 52.9 and   62.5 GeV,
that were already analyzed in ref.~\cite{Bialas:2006qf}, 
new elastic scattering data became available recently 
at $\sqrt{s} = 7$ TeV~\cite{Antchev:2011zz}
from the CERN LHC experiment TOTEM. The new data triggered a wide spectrum of theoretical investigations, 
including Regge theory based studies~\cite{Fagundes:2011wn, Martin:2011gi, Donnachie:2011aa, Fagundes:2011zx} involving perturbative 
QCD(BFKL)~\cite{Ryskin:2012ry}, as well as study with hidden dimensions~\cite{Block:2012yx}. Extrapolation to the domain of cosmic rays~\cite{Wibig:2011iw} become available.
We have tested both variations of the model of Bialas and Bzdak of ref.~\cite{Bialas:2006qf}, 
not only at ISR energies but also at the currently available
highest LHC energies on recent TOTEM data,
in order to learn more details about the evolution of the properties 
of p+p elastic interactions in the recently opened, few TeV energy range. The
two variations of the Bialas and Bzdak model allowed us to obtain an effective
radius parameter $R_{\mbox{\rm\scriptsize eff}}$ which is model independent within errors. A simple relation between the total cross section
and the effective radius is also presented in the discussions.

		Most of the arguments for the composite structure of hadrons have been 
derived from the studies of lepton-hadron interactions. The standard
picture is that hadrons are either mesons, composed of valence quarks and anti-quarks,
or (anti)baryons, composed of three valence (anti)quarks, that
carry the quantum numbers, while the electrically neutral gluons
carry color charges and provide the binding among the quarks and anti-quarks. 
The exact contribution of quarks, gluons, and the sea of virtual quark-antiquark pairs and gluons
to certain hadronic properties e.g. spin is still under detailed investigation.
Also, more than 10 exotic hadronic resonances called X, Y and Z states were recently discovered
in electron-positron collisions at the world's highest luminosities in the 
BELLE experiment at KEK. These hadronic states 
cannot be interpreted in the standard picture of quark-antiquark 
or three (anti)quark bound states, according to refs.~\cite{KEK-XYZ,Kreps:2009ne}.
Thus even nowadays there are still several open questions that are related to the
compositeness of the hadrons in general.
As gluons do not interact directly with leptons, their properties are best explored
with the help of the strong hadronic interactions. For example, the gluon contribution
to the proton spin is investigated with the help of  polarized proton - polarized proton collisions
at RHIC~\cite{Adare:2008px}, but it is still not fully constrained.
In this work, we focus on the effects of internal correlations between the quarks
inside the protons, reporting on a detailed study of proton-proton elastic scattering 
at several ISR and also at the currently top available LHC energies.
Let us recall, that a similar analysis involving three {\it independent} quarks 
was not able to properly describe the ISR data ~\cite{Bialas:1977xp}.
In that model quarks were considered as ``dressed'' valence quarks 
in the sense that they contain the gluonic and $q\bar{q}$ contribution as well, as if 
the glue would be concentrated around pointlike valence quarks. 
Thirty-five years after the three independent quark model of ref.~\cite{Bialas:1977xp}
another three-quark model of the protons was proposed in ref.~\cite{Bialas:2006qf}, 
that included interesting correlations between 
two dressed valence quarks to form a diquark. In the present investigation this quark-diquark 
model~\cite{Bialas:2006qf}, is compared to data in details.  

The quark-diquark  picture of elastic p+p scattering 
is similar to the Glauber optical model~\cite{Glauber:1955qq} 
in nuclear physics.
This Glauber model, developed originally for nuclear multiple scattering 
problems like cross sections of protons and neutrons on deuteron, 
became a standard model of high energy interactions in nuclear physics
where multiple interactions are built up from superpositions
of nucleon-nucleon scattering. This model became a fundamental and
successful tool in describing nuclear collisions at high energy ~\cite{Glauber:2006gd}. 

\par

   The body of this manuscript is organized as follows: 
   the theoretical model  of Bialas and Bzdak is recapitulated and summarized in Section \ref{sec:theory}, 
   including two cases. In the first case, the proton is modelled as a quark-diquark composite system,
   while in the second case, the diquark is assumed to have a quark-quark internal structure.
   Section \ref{sec:minuit}
   contains our final results, which is based on a multiparameter optimalization procedure,
   utilizing the CERN MINUIT package \cite{James:1975dr}. Finally, we summarize and conclude in Section \ref{sec:conclusion}.

\section{Elastic scattering in the quark-diquark model}
\label{sec:theory}
\begin{figure}[H]
    	\includegraphics[width=0.6\textwidth]{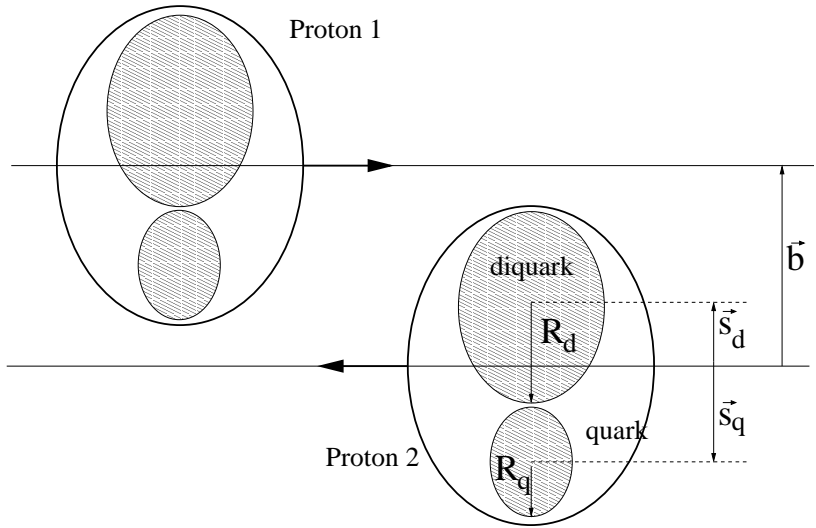}
		\centering
		\caption{Scheme of the scattering of two protons, when the proton is assumed to have a quark-diquark
        structure and the diquark is assumed to be scattered as a single entity. This is just a snapshot and 
		all the model parameters follow a Gaussian distribution. Note, that a center of mass energy 
        dependent Lorentz-contraction determines the longitudinal scale parameters.}
		\label{scattering sit1}
    \end{figure}

	We describe proton-proton interactions as collision of two systems, each one composed of a dressed quark and diquark. The 
	p+p collision is schematically illustrated
	on Fig. \ref{scattering sit1}.

	The interaction between quarks and diquarks is assumed to be purely absorptive. Consequently the amplitude has no real part and the imaginary part -- 
	dominating at high energy -- is given by the absorption of the incoming particle wave, represented by the inelastic (non-diffractive) collisions.\par
    In the impact parameter space the inelastic proton-proton cross-section for a fixed impact parameter $\vec{b}$ can be given by the following formula \cite{Bialas:2006qf}
	\begin{equation}
		\sigma(\vec{b})=\int\limits^{+\infty}_{-\infty}...\int\limits^{+\infty}_{-\infty}{\text{d}^2s_q \text{d}^2s'_q \text{d}^2s_d \text{d}^2s'_d D(\vec{s_q},\vec{s_d})
		D(\vec{s_q}',\vec{s_d}')
		\sigma(\vec{s_q},\vec{s_d};\vec{s_q}',\vec{s_d}';\vec{b})},
		\label{elsoegyenlet}
	\end{equation}
	where $\vec{s_{q}}$, $\vec{s_{q}}'$ and $\vec{s_{d}}$, $\vec{s_{d}}'$ are the transverse positions of 
	the quarks and diquarks respectively. The integrand is a product of quark-diquark distributions of the
        incoming protons and a $\sigma$ function which gives the probability of inelastic interaction at given 
        impact parameter $\vec{b}$ and at given quark, diquark positions.\par 
	The quark-diquark distribution inside the nucleon is taken into account with the following Gaussian
    \begin{equation}
    	D\left(\vec{s_q},\vec{s_d}\right)=\frac{1+\lambda^2}{\pi R_{qd}^2}e^{-(s_q^2+s_d^2)/R_{qd}^2}\delta^2(\vec{s_d}+\lambda \vec{s_q}),\;\lambda=m_q / m_d,
    \end{equation}
	where  $R_{qd}$ is the RMS of the separation between the CMS of the diquark and the remaining quark in the proton. $\lambda$ is the mass 
	ratio of the quark and the diquark. Obviously $ 1/2 \le \lambda \le 1$, 
	where $1/2$ would indicate a loosely bound diquark. The two dimensional delta function fixes the position of the center of mass in the transverse plane.\par
	Elastic interactions are {\it independent}, accordingly the probability distribution of elastic 
        proton-proton collision is the {\it product} of the probability distribution of elastic interactions 
	of their constituents \cite{Glauber,Czyz:1969jg}
	\begin{align}
		\sigma(\vec{s_q},\vec{s_d};\vec{s_q}',\vec{s_d}';\vec{b})=1-\prod_{a,b \in \{q,d\}}\left[1-\sigma_{ab}(\vec{b} + \vec{s_a}' - \vec{s_b}' )\right].
	\label{Glauber expansion}
	\end{align}
	The inelastic differential cross-sections are parametrized with Gaussian distributions
	\begin{equation}
		\sigma_{ab}\left(\vec{s}\right) = A_{ab}e^{-s^2/R_{ab}^2},\;R_{ab}^2=R_a^2+R_b^2,
		\label{inelastic cross sections}
	\end{equation}
	where $R_{ab}$ is the variance of having an inelastic collision, which is calculated from the sum of the squared $R_q$, $R_d$ radius parameters; the $A_{ab}$ parameters
	are the amplitudes. From unitarity the elastic amplitude in impact parameter space
		\begin{equation}
			t_{el}(\vec{b})=1-\sqrt{1-\sigma(\vec{b})}.
		\end{equation}
	As it was mentioned the real part of the amplitude is ignored. Recently, the important role of the real part of
    the elastic scattering amplitude in shaping $d\sigma/dt$ at the dip and in the Orear region was highlighted in \cite{Dremin:2012dm, Dremin:2012yh}. \par
	The elastic amplitude in momentum transfer representation is the Fourier-transform of the amplitude in impact parameter space 
		\begin{equation}
			T(\vec{\Delta})=\int\limits^{+\infty}_{-\infty}\int\limits^{+\infty}_{-\infty}{t_{el}(\vec{b})e^{i\vec{\Delta} \cdot \vec{b}}\text{d}^2b}=
			2\pi\int\limits_0^{+\infty}{t_{el}\left(b\right)J_0\left(\Delta b\right)b {\text d}b},
		\end{equation}
	where $\Delta=|\vec{\Delta}|$, $b=|\vec{b}|$ and $J_0$ is the zeroth Bessel-function of the first kind. Then the elastic differential cross section reads as
	\begin{equation}
		\frac{d\sigma}{dt}=\frac{1}{4\pi}\left|T\left(\Delta\right)\right|^2.
	\end{equation}


\subsection{Model $p=(q,d)$: The diquark is assumed to scatter as a single entity}
	
	The subject of this section is to analyze, as the first investigated case,
    that variant of the model of Bialas and Bzdak, when the quark and the 
    diquark is considered to scatter as one entity as
	indicated on Fig. \ref{scattering sit1}. In this case, the number of free parameters 
    can be reduced if we assume that the number of partons is twice as many in the 
    diquark than in the quark. From the inelastic differential 
    cross sections (\ref{inelastic cross sections}) the total inelastic cross sections are
	\begin{equation}
	\label{totalinelastic}
		\sigma_{ab}=\int\limits^{+\infty}_{-\infty}\int\limits^{+\infty}_{-\infty}{\sigma_{ab}\left(\vec{s}\right)}\text{d}^2s= \pi A_{ab}R_{ab}^2,\;\; a,b \in \{q,d\}.
	\end{equation}
	Our assumption tells us that 
	\begin{equation}
		\sigma_{qq}:\sigma_{qd}:\sigma_{dd}=1:2:4, 
		\label{ratiosforsigma}
	\end{equation}
	from which we can deduce the following expressions
		\begin{equation}
			A_{qd}=A_{qq}\frac{4R_q^2}{R_q^2+R_d^2},\;A_{dd}=A_{qq}\frac{4R_q^2}{R_d^2},
		\end{equation}
	which means that every $A_{ab}$ parameter can be expressed in term of $A_{qq}$. With these ingredients the calculation of each term in (\ref{elsoegyenlet}) reduces
	to Gaussian integrations. Two of the Dirac $\delta$ functions in (\ref{elsoegyenlet}) induce the following transformation in the transverse
    diquark and quark position variables
    \begin{equation}
        \vec{s_d}=-\lambda \vec{s_q},\, \vec{s_d}'=-\lambda \vec{s_q}'.
    \end{equation}
    Hence four Gaussian integration remain, which lead us to the following result \cite{Bialas:2006qf}
    \begin{align}
			\label{master_formula}
        &\frac{4v^2}{\pi^2}\int\limits^{+\infty}_{-\infty}\int\limits^{+\infty}_{-\infty}{\text{d}^2s_q \text{d}^2s_q' e^{-2v\left(s_q^2+s_q'^2\right)}
            e^{-c_{qq}\left(b-s_q+s_q'\right)^2} e^{-c_{qd}\left(b-s_q+s_d'\right)^2}} \\
            &\times e^{-c_{dq}\left(b-s_d+s_q'\right)^2} e^{-c_{dd}\left(b-s_d+s_d'\right)^2}=\frac{4v^2}{\Omega}e^{-b^2\frac{\Gamma}{\Omega}} \notag,
    \end{align}
    where the coefficients $c_{ab}$ are abbrevations, and
    \begin{align}
        \Omega=&\left[4v + \left(1+\lambda\right)^2\left(c_{qd}+c_{dq}\right) \right]\left[v+c_{qq} + \lambda^2 c_{dd}\right]\\
        &+\left(1-\lambda\right)^2\left[v \left(c_{qd}+c_{dq}\right)+\left(1+\lambda\right)^2c_{qd}c_{dq}\right] \notag,
    \end{align}
    while
    \begin{align}
        \Gamma=&\left[4v + \left(1+\lambda\right)^2\left(c_{qd}+c_{dq}\right) \right]\left[v \left(c_{qq}+c_{dd}\right)+\left(1+\lambda\right)^2c_{qq}c_{dd}\right]\\
        &+\left[4v + \left(1+\lambda\right)^2\left(c_{qq}+c_{dd}\right) \right]\left[v \left(c_{qd}+c_{dq}\right)+\left(1+\lambda\right)^2c_{qd}c_{dq}\right] \notag.
    \end{align}

\subsection{Model $p= (q,(q,q))$: The diquark is assumed to scatter as a composite object}
\begin{figure}[H]
        \includegraphics[width=0.6\textwidth]{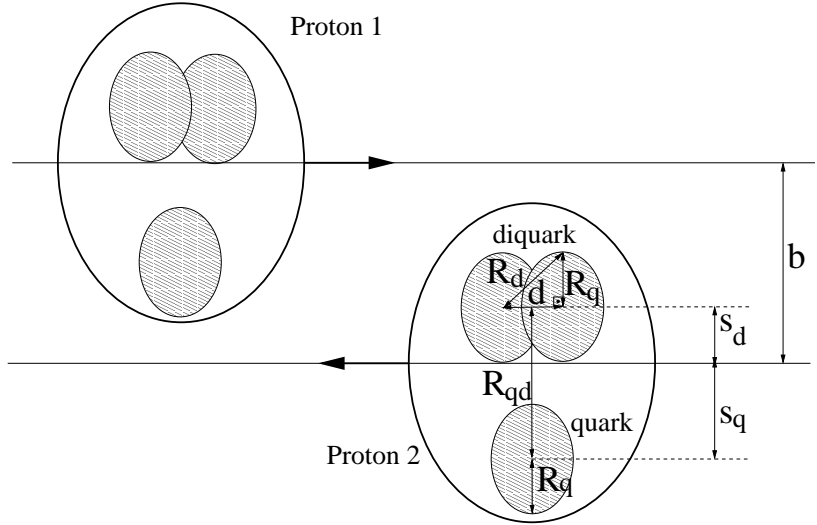}
        \centering
        \caption{The scattering situation of the two protons when the diquark is assumed to be composed of two quarks, and the proton symbolically can be written as p=(q,(qq)).
		This is just a snapshot and all the model parameters follow a Gaussian distribution.}
        \label{scatter qq}
    \end{figure}

		When the diquark is assumed to scatter as a composite object, that includes two  valence quarks, 
        the scheme of elastic p+p scattering is illustrated on Fig. \ref{scatter qq}.  
        Following Bialas and Bzdak~\cite{Bialas:2006qf},  the quark distribution inside the diquark is supposed
        to have the following Gaussian shape
    \begin{equation}
            D\left(\vec{s_{q1}},\vec{s_{q2}}\right)=\frac{1}{\pi d^2}e^{-\left(s_{q1}^2+s_{q2}^2\right)/2d^2}
        \delta^2\left(\vec{s_{q1}}+\vec{s_{q2}}\right),
        \label{quarkdistribution}
    \end{equation}
    where $\vec{s_{q1}}$ and $\vec{s_{q2}}$ are the transverse quark positions inside the diquark, and
    \begin{equation}
        d^2=R_d^2-R_q^2
    \end{equation}
    is the RMS of the separation of quarks inside the diquark, calculated from the diquark and quark radius parameters.
        If the internal structure of the diquark is given by eq. (\ref{quarkdistribution}) then the $\sigma_{qd}$,
    $\sigma_{dq}$ and $\sigma_{dd}$ inelastic differential cross sections eq. (\ref{inelastic cross sections})
    can be calculated from $\sigma_{qq}$ using an expansion analogous to expression eq. (\ref{Glauber expansion}). The result
    for $\sigma_{qd}$ is the following
    \begin{equation}
            \sigma_{qd}\left(\vec{s}\right)=\frac{4A_{qq}R^2_q}{R^2_d+R^2_q}e^{-s^2\frac{1}{R^2_d+R^2_q}}-\frac{A^2_{qq}R^2_{q}}{R^2_d}e^{-s^2/R^2_q},
    \end{equation}
     $\sigma_{dd}$ is a bit more complicated~\cite{Bialas:2006qf}. The inelastic cross section of eq.~(\ref{elsoegyenlet}) for this model is again obtained using expression~(\ref{master_formula}).\par

  In their original paper, Bialas and Bzdak~\cite{Bialas:2006qf} 
  fixed the model parameters
  by the total cross section, the slope parameter $B$, the position of the dip and the 
  position of the first diffractive maximum after the dip. This way, they determined
  the best values of the model parameters essentially by the method of numerical solution
  of four equations containing four parameters, and found, that the resulting solution
  gives a good overall description of elastic scattering data at ISR energies.
    
    In the first phase of our studies, we confirmed their results, and noted that if we want to
    determine the errors on the model parameters a different strategy is needed.
    As we were interested in finding significant dependence of the model parameters
    on the colliding energy, for which the estimation of the errors on the model parameters
    is essential, we started to improve on the original method of Bialas and Bzdak 
    by a multiparameter fit utilizing all the available information in the data set, determining the best
    confidence levels and the errors on the model parameters.

  In the next section,  we present the final results of this multi-parameter optimalization method.
  This manner, we improve on the earlier analysis of Bialas and Bzdak~\cite{Bialas:2006qf}
  with the help of the CERN Minuit package, in both cases: when the diquark acts as a 
  single object, and also when the diquark is considered as an object composed of two quarks.
  We perform these analysis for all the energies where  p+p elastic scattering data are available from the
  ISR collider, and we apply the same kind of analysis
  also to recent TOTEM data at $\sqrt{s} = 7 $ TeV at CERN LHC energy.

  Our final analysis results of p+p elastic scattering data  are presented
  for data sets that are restricted to exactly the same kinematic domain 
  at all the five considered energies.  We also study, 
  in addition to the differential cross-section
  of elastic scattering, the effect of including the measured total p+p scattering 
  cross-section data to the fitting procedure,  at all the four ISR  and also at $\sqrt{s} = 7 $ TeV LHC energies.

  Finally, we summarize and conclude.

\newpage

\section{Final MINUIT fit results to p+p elastic scattering data at ISR and LHC energies}
\label{sec:minuit}

	\subsection{Model $p=(q,d)$: The diquark is assumed to scatter as a single entity}

	In this section the MINUIT fit results are presented for the 
    ISR \cite{Nagy:1978iw,Amaldi:1979kd} and TOTEM~\cite{Antchev:2011zz} proton-proton elastic scattering 
	data considering the scenario when the diquark is assumed to act as a single entity in this scattering process.
    Our preliminary analysis of these data, that did not yet study the correlations between model parameters
    and also did not evaluate what happens if some of the model parameters are fixed, was 
    presented in a recent conference contribution ~\cite{Nemes:2012mq}.

\begin{figure}[H]
		\includegraphics[width=0.9\textwidth]{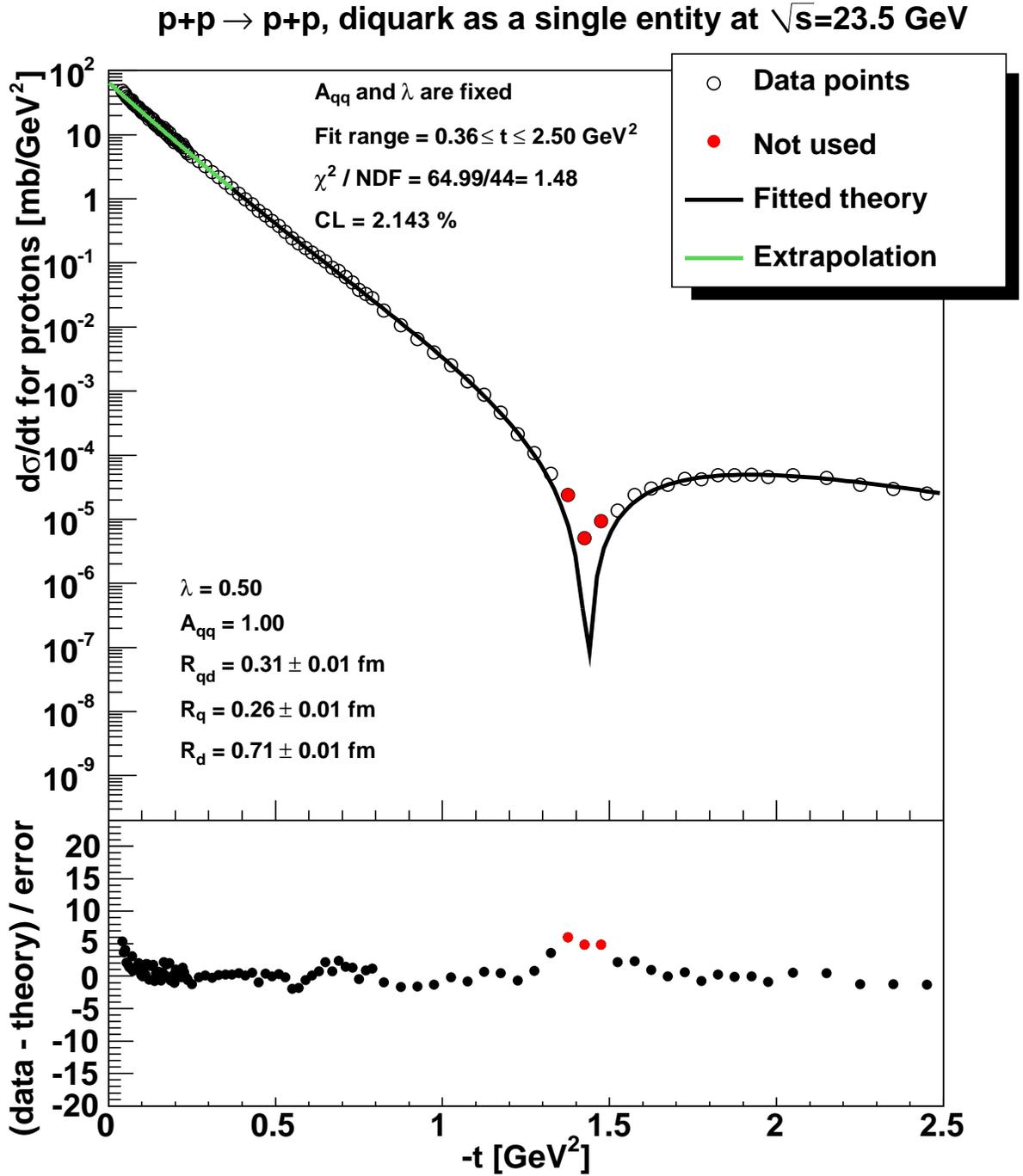}
        	\centering
                \caption{(Color online.) Results of MINUIT fits at ISR energies when the 
                diquark is assumed to scatter as a single entity.  Top panel shows the data points
                and the result of the best fit, while the lower panel shows the relative deviation of the model
                from data in units of measured error bars.
				As the model is singular around the dip, 3 data points that are located closest to this diffractive 
                minimum and are indicated with filled (red) circles in this Figure, were left out from the fit.
				The optimalization was restricted to the 0.36 - 2.5 GeV $\left|t\right|$ range,
                so that a fair comparison could subsequently be made with most recent TOTEM results
                as given in ref.~\cite{Antchev:2011zz}. 
                The best fit is indicated with a solid (black) line in this range and its extrapolation
                to low values of $t$ are also shown.    
                The confidence level, after fixing the values of $\lambda$ and $A_{qq}$,
                is still higher than 0.1\%, which indicates that this fit quality is statistically acceptable.  }

		\label{singlefitfor23}
	\end{figure}

\begin{figure}[H]
					\includegraphics[width=0.9\textwidth]{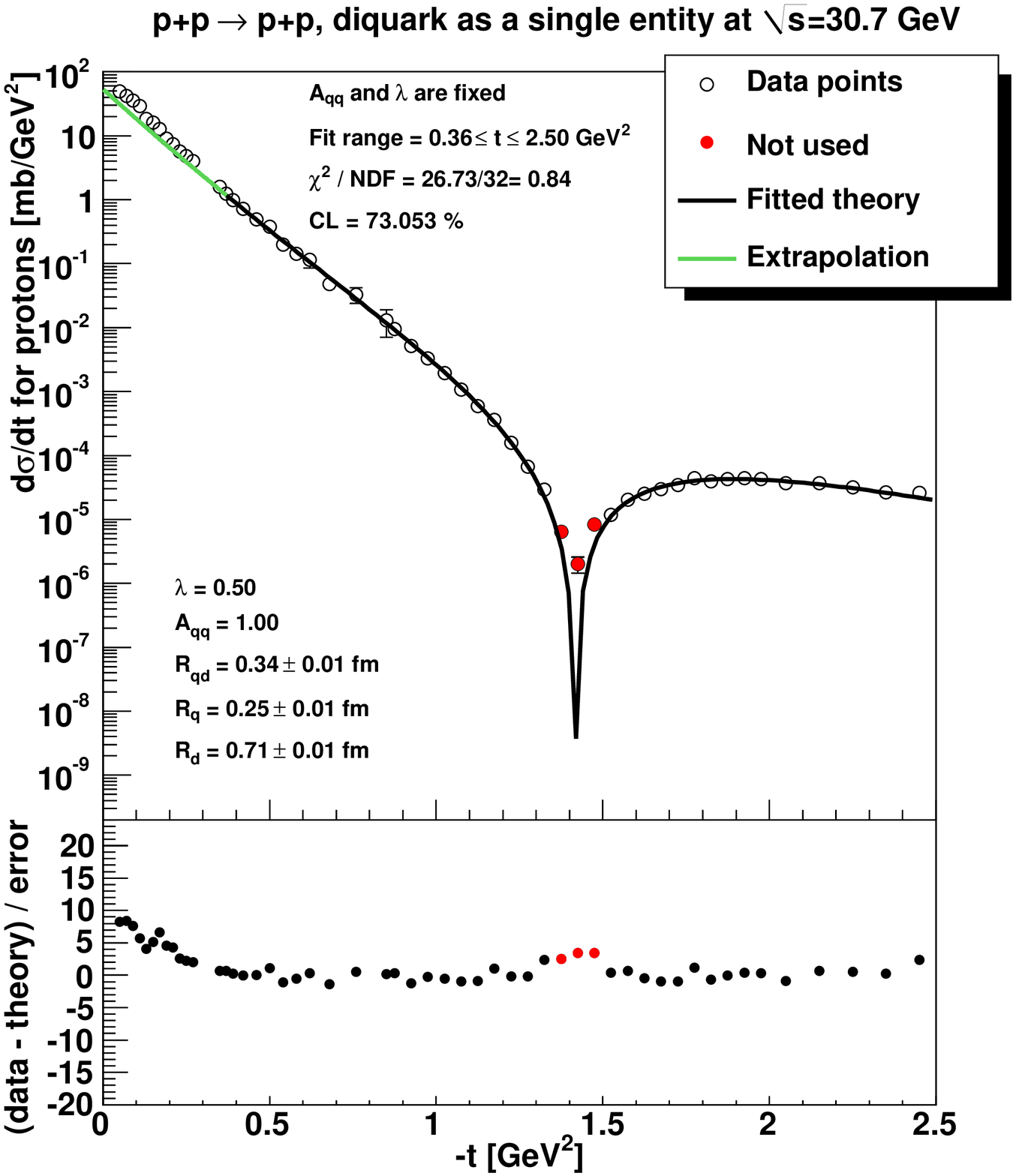}
                    \centering
                    \caption{(Color online.) Same as Fig. \ref{singlefitfor23}, but for the energy $\sqrt{s}=30.7$ GeV. }
			\label{singlefitfor31}
                \end{figure}

\begin{figure}[H]
					\includegraphics[width=0.9\textwidth]{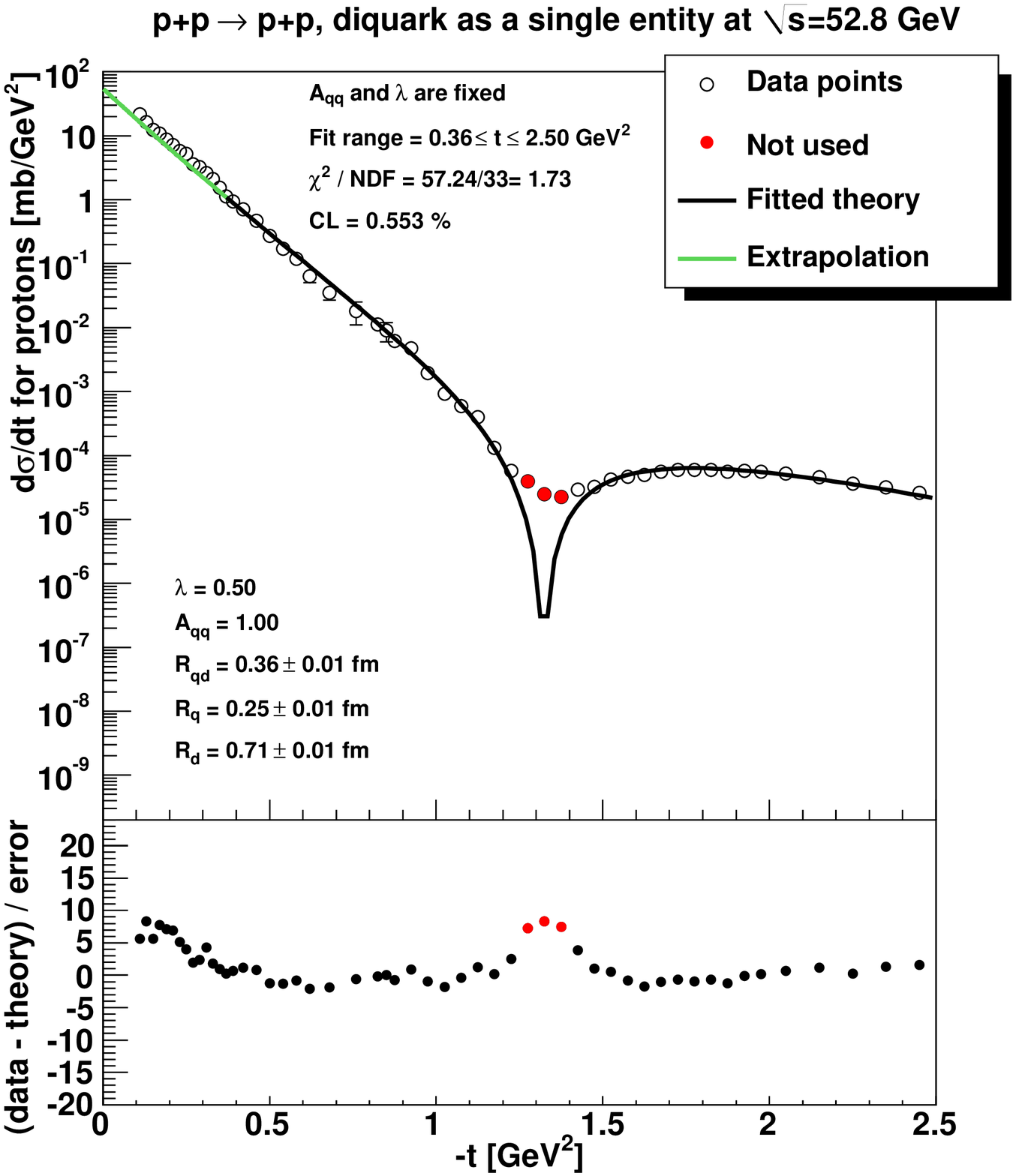}
                    \centering
                    \caption{(Color online.) Same as Fig. \ref{singlefitfor23}, but for the energy $\sqrt{s}=52.8$ GeV. }
			\label{singlefitfor53}
	\end{figure}

\begin{figure}[H]
			\includegraphics[width=0.9\textwidth]{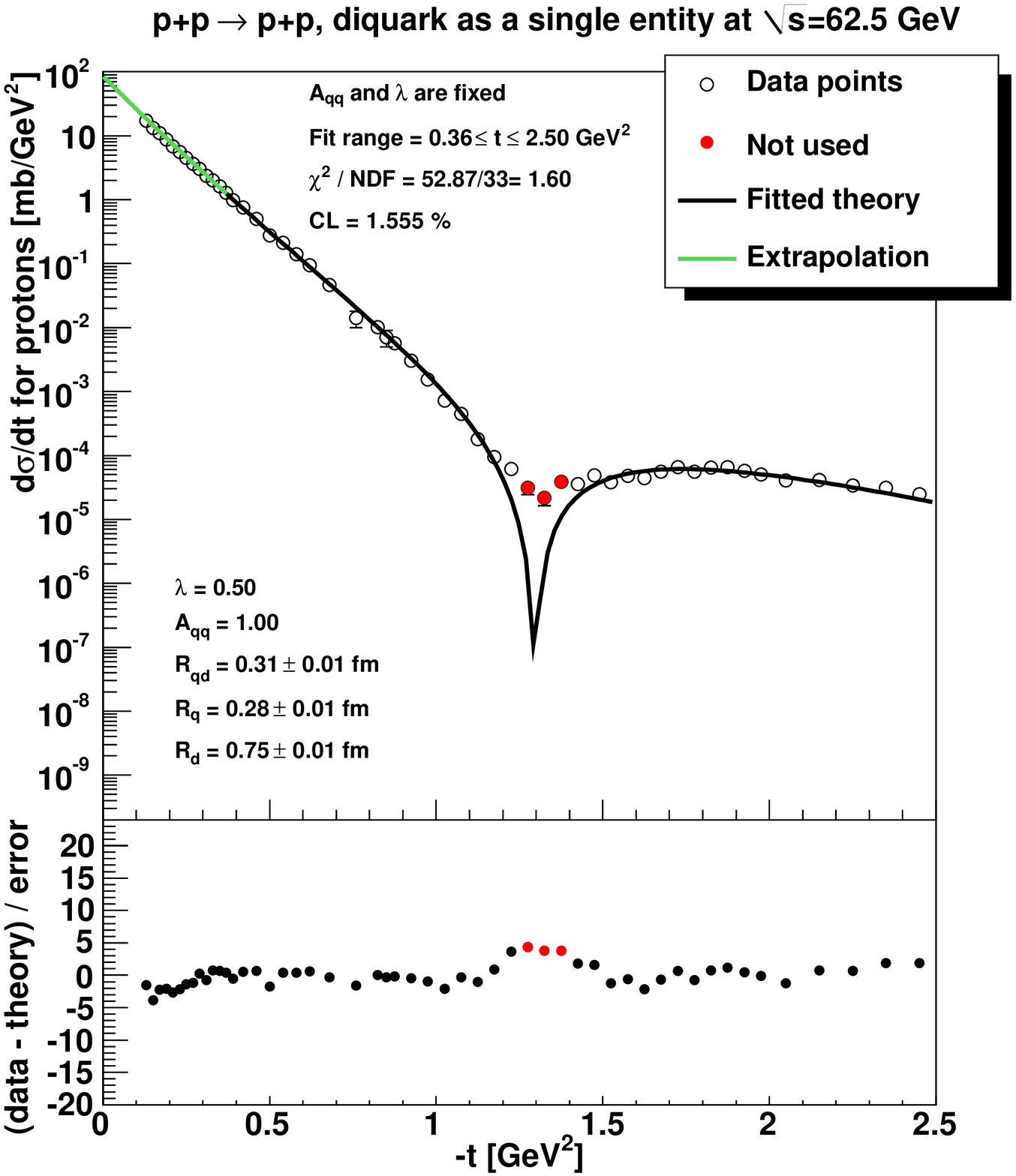}
			\centering
			\caption{(Color online.) Same as Fig. \ref{singlefitfor23}, but for the energy $\sqrt{s}=62.5$ GeV.}
		\label{singlefitfor62}
	\end{figure}

\begin{figure}[H]
                \includegraphics[width=0.9\textwidth]{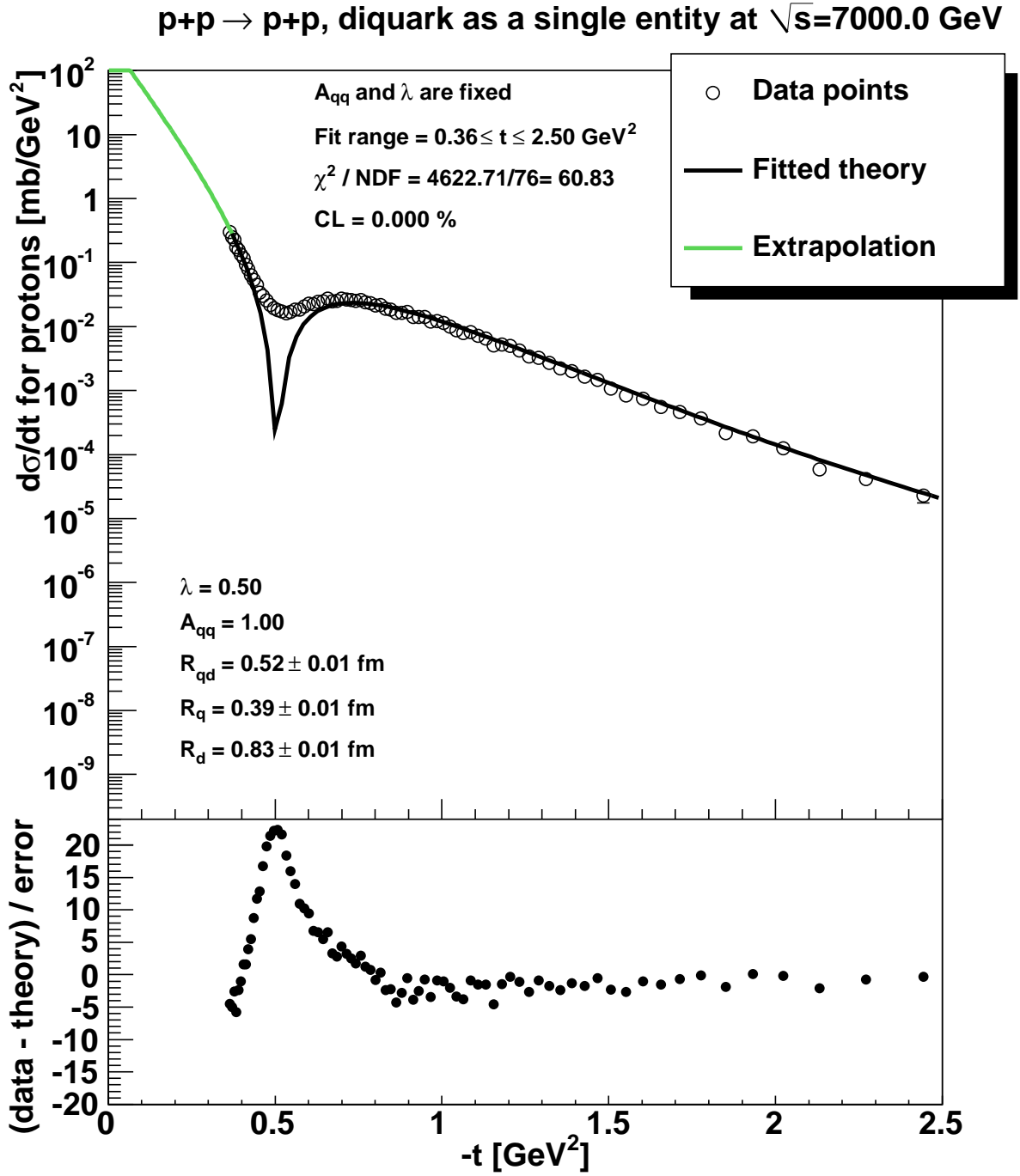}
                \centering
		\caption{(Color online.) The result of the fit at LHC at $7$ TeV when the diquark is assumed to scatter as a single entity,
        $p = (q, d)$. Note that CL is below 0.1\%, so the quality of this fit is not acceptable.
        The bottom panel indicates, that the shape of the diffractive cross-section around the
        first diffractive minimun is not reproduced correctly by this model at LHC energies, and -- as can also be seen on
	this Figure -- this shortcoming
        cannot be fixed by leaving out a few data points around this dip from the  optimalization procedure.}
		\label{singlefittotems}

	\end{figure}

    Our final results are illustrated on Figs. \ref{singlefitfor23}-\ref{singlefittotems}. 
    The confidence levels, and model parameters together with their errors are presented 
    in Table \ref{parameters single}. The calculated total elastic cross sections,
    including their uncertainty, were evaluated from the MINUIT fits to the differential
    cross section data, restricted to the same intermediate $t$ elastic scattering.
    At the end of this section we also study and discuss, what happens when the measured total elastic
    cross sections are also added, as an additional data point, to the optimalization procedure.

    The ratios of the inelastic cross sections were fixed with eq. (\ref{ratiosforsigma}),
    in order to decrease the number of free fit parameters, 
    therefore these ratios will be provided only for the case,
    when the diquark is assumed to be a composite object.\par  

    Some additional remarks are due before our final results are presented. 
    The Bialas - Bzdak model shows a singular behaviour at the diffractive minimum or dip position,
    which is due to the lack of a real part in its amplitude. 
    In a Minuit fit, such an unphysical fit region may completely dominate the fit results.
    In order to avoid such an artefact and to obtain a meaningful fit
    result, we have excluded 3 data points from the optimalization procedure,
	that were closest to this dip region. These points are shown in red (color online) or on the plots. 
	We have checked that leaving out 5 or 7 points did not change the results.\par 
	Another important remark is that the TOTEM data covers the $|t|$ range from 0.36 GeV up to 2.5 GeV 
    and this range is applied
    in our minimization procedure to allow a comparison between the ISR and TOTEM results. 
	Note that Bialas and Bzdak adjusted their model in a different way:
    they demanded, that four important features of the data are described correctly,
    namely  (i) the total inelastic cross-section and (ii) the slope 
    of the differential cross-section as extrapolated to $t = 0$, and (iii) the position of the 
    first diffractive  minimum and finally (iv) the height of the diffractive maximum 
    after the diffractive minimum ~\cite{Bialas:2006qf}.  
    A different strategy is followed here, since we have fitted  the theoretical curve directly 
    to the experimental data points using the CERN MINUIT package by minimizing $\chi^2$. However, we selected the same
    reasonable $t$-range for all data sets, and we left out data points around the diffractive
    minimum, where this model clearly breaks down. These data points are clearly indicated with filled 
    (color online red) circles in the subsequent Figures, where the fit results are shown and can  
    be compared to data directly.\par
    In a preliminary analysis, reported in ref.~\cite{Nemes:2012mq}, all model parameters
    were optimized, including also the values of $\lambda$ and    $A_{qq}$.
    Those preliminary results are compatible with the final fits presented here, however,
    in those results large correlations remained among the fit parameters, and their errors
    were correspondingly large.
    The optimalization process indicated that the parameters
    $\lambda $ and $A_{qq}$ were in the range of their nominal value and fixing them resulted in
    considerable reduction of  correlations among the other model parameters. Hence in the
    final fits presented here we utilized fixed values of $\lambda = 0.5 $ and $A_{qq} = 1.0$.

\vfill
\newpage

				\begin{figure}[H]
                	\includegraphics[trim = 6mm 12mm 2mm 4mm, clip, width=0.45\textwidth]{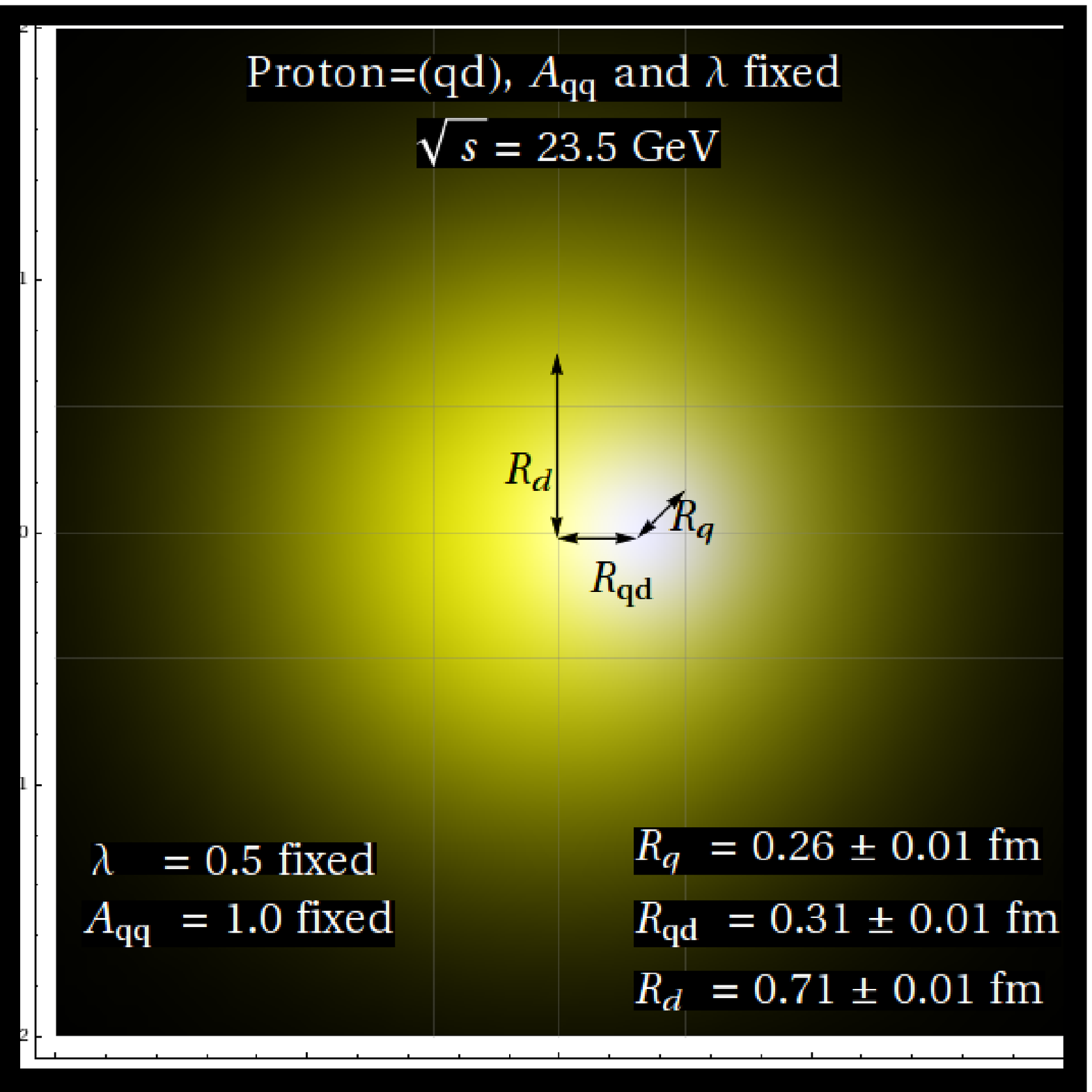} 
                	\includegraphics[trim = 6mm 12mm 2mm 4mm, clip,width=0.45\textwidth]{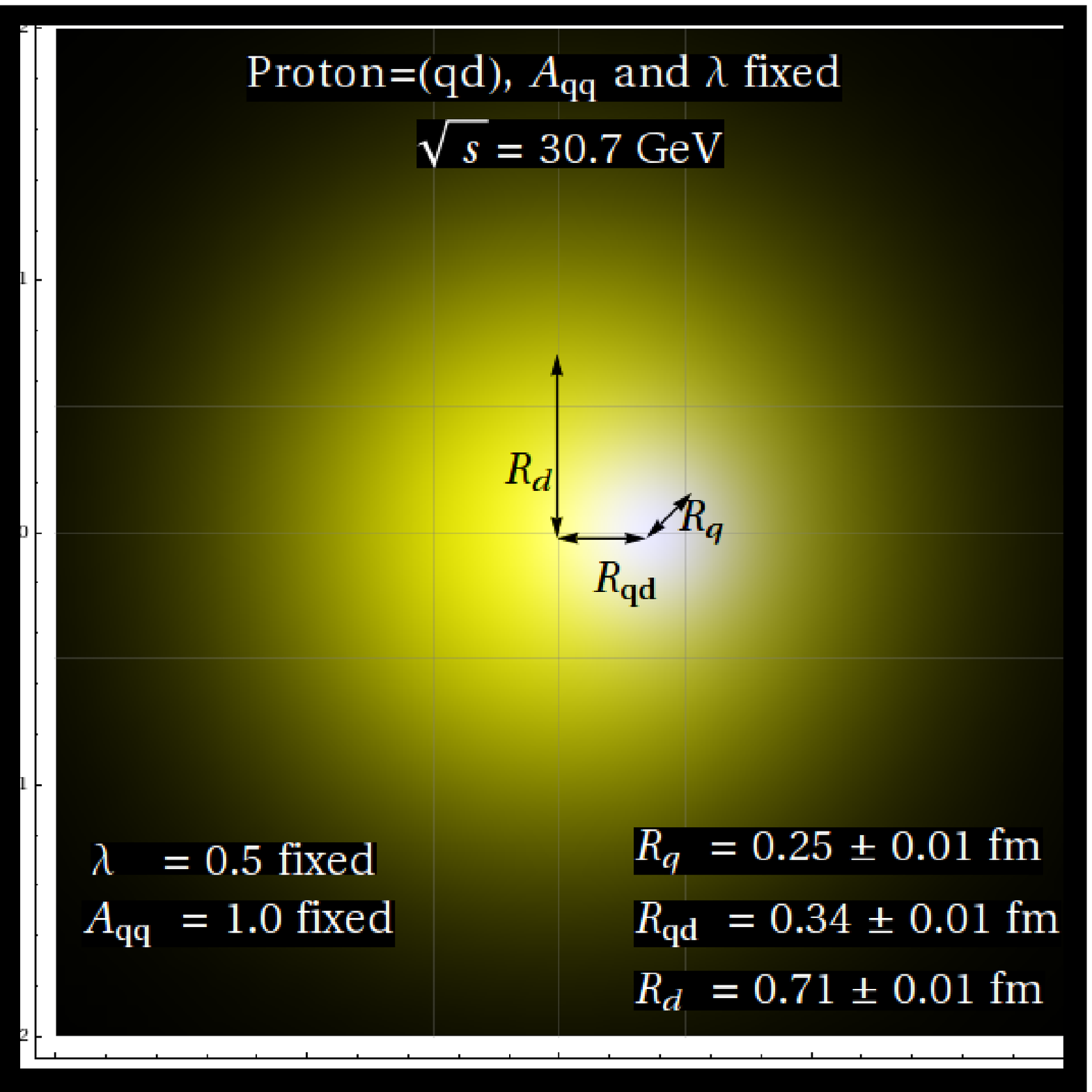} 
                	\includegraphics[trim = 6mm 12mm 2mm 4mm, clip,width=0.45\textwidth]{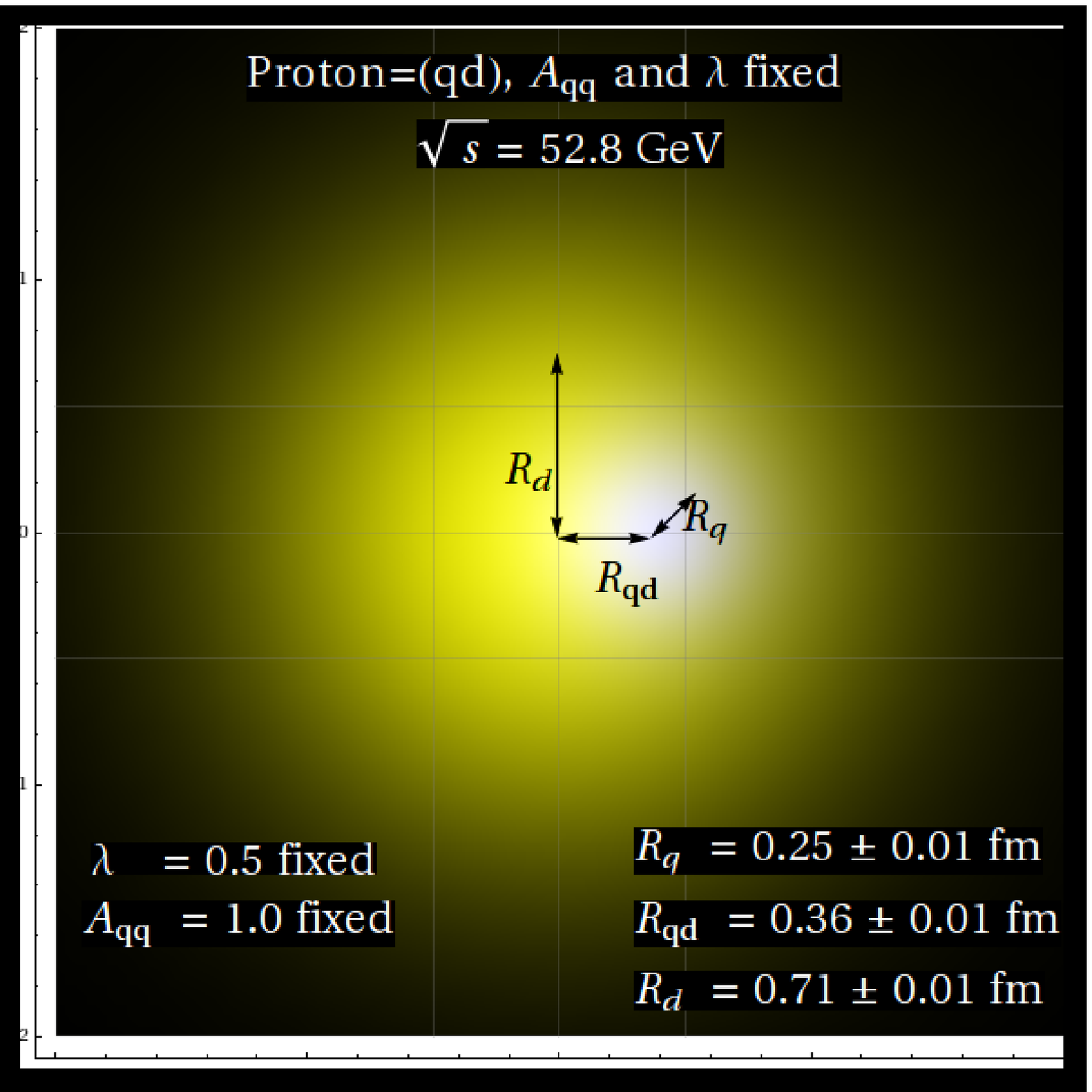} 
                	\includegraphics[trim = 6mm 12mm 2mm 4mm, clip,width=0.45\textwidth]{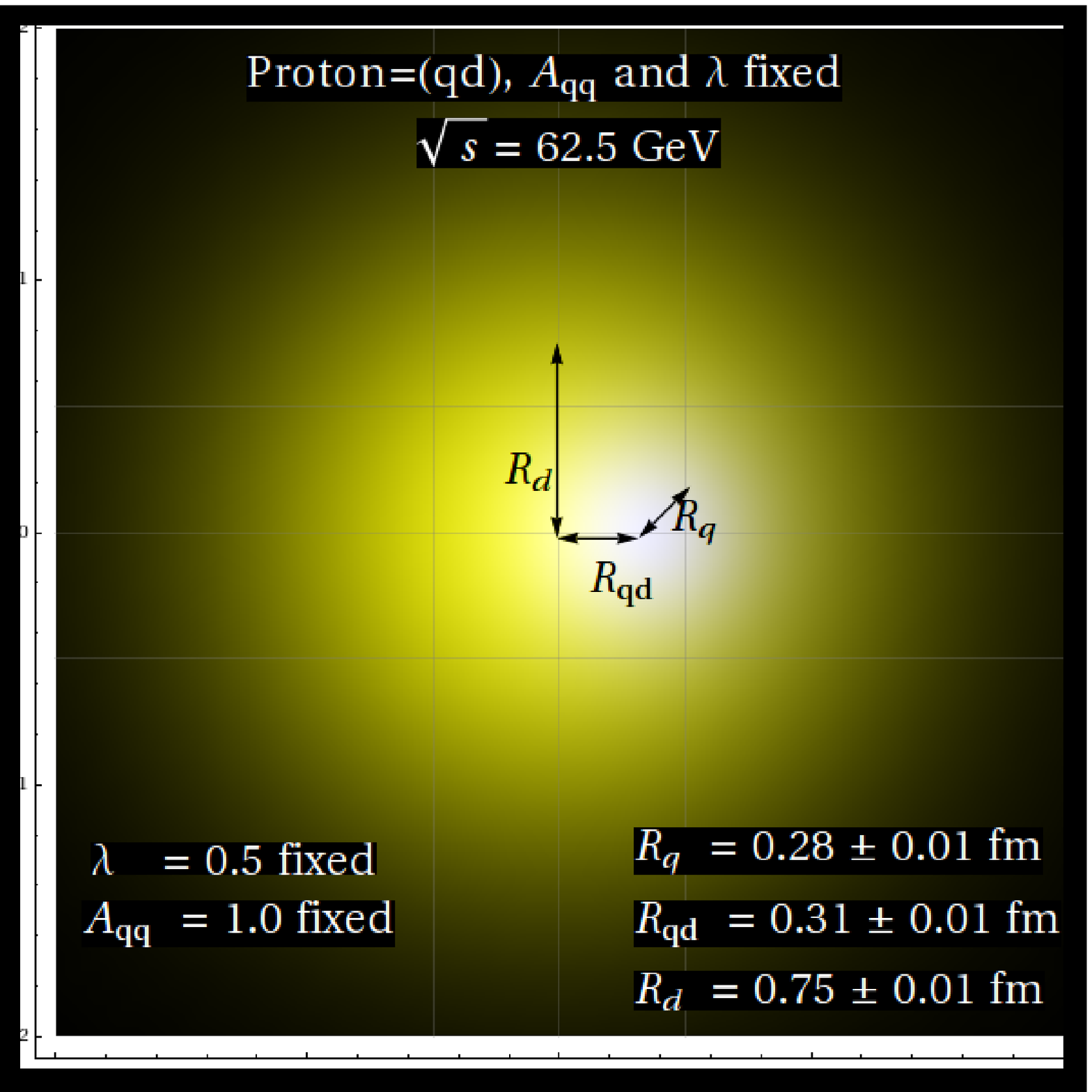} 
                	\includegraphics[trim = 6mm 12mm 2mm 4mm, clip,width=0.45\textwidth]{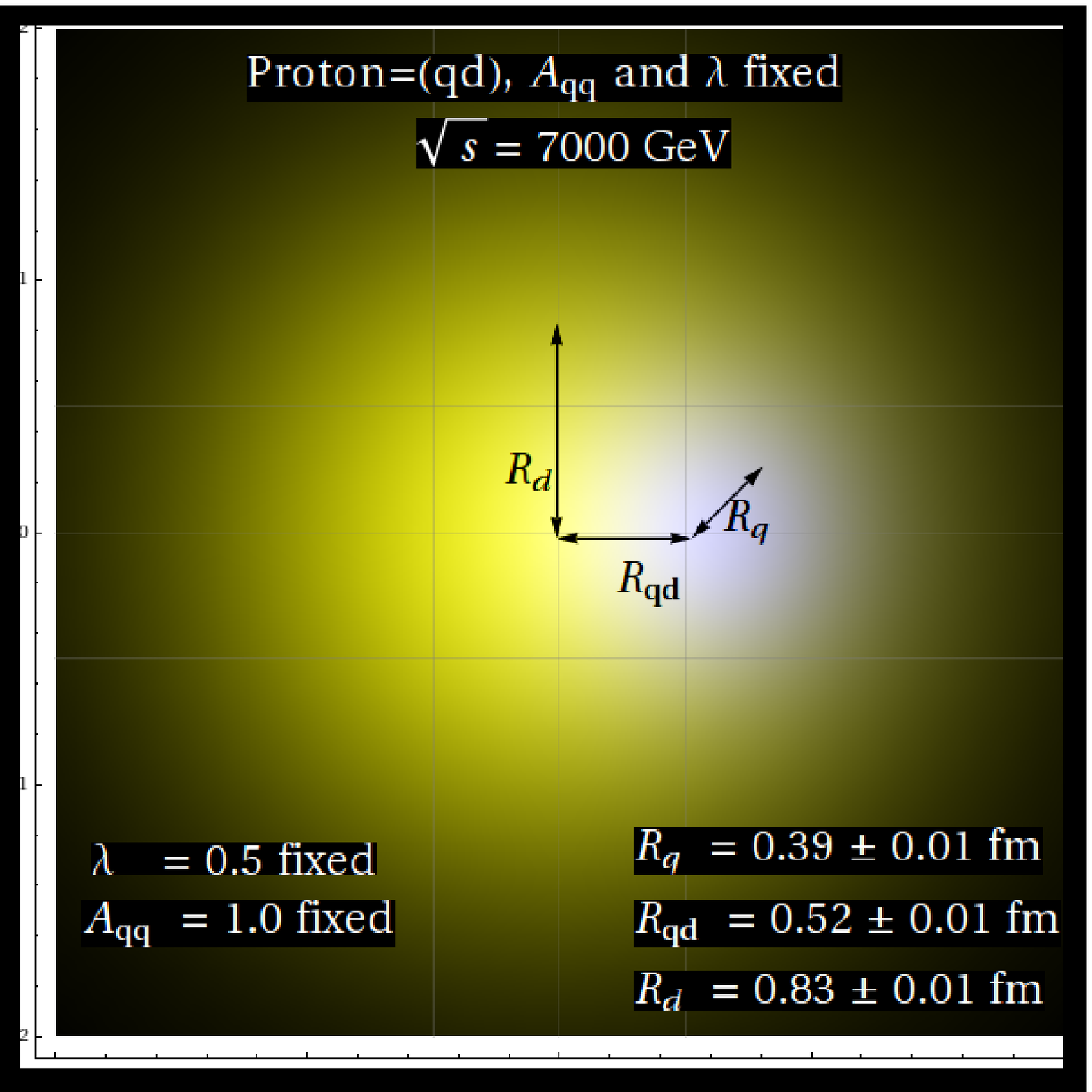} 
		\centering
		\caption{(Color online.) Visualisation of the obtained $R_{qd}$, $R_q$, $R_d$ parameters for the 
        case when the proton is assumed to scatter as  a quark-diquark composit, $p= (q, d)$
        and $A_{qq} = 1$, $\lambda = 0.5 $ fixed. The main observation is that the 
			 proton seems to be much larger at LHC energies than at ISR regime. 
			This is mainly due to an increase in the $R_{qd}$ parameter,
			that characterizes the separation of the quark and the diquark.
                    }
		\label{single visualisation}
		\end{figure}

	\begin{table}\footnotesize
	    \begin{tabular}{|c|c|c|c|c|c|} \hline
			$ \sqrt{s}$ [GeV]	& 23.5  			& 30.7 				&  52.8				& 62.5 				& 7000   \\ \hline\hline 
			$\lambda$ 			& 0.5 	 			&  0.5  			&  0.5  			&  0.5  			&  0.5  \\ \hline 
			$A_{qq}$ 			& 1.00  			&  1.00  			&  1.00  			&  1.00  			&  1.00  \\ \hline 
			$R_{qd}$ [fm] 		& 0.31 $\pm$ 0.01 	&  0.34 $\pm$ 0.01 	&  0.36 $\pm$ 0.01 	&  0.31 $\pm$ 0.01 	&  0.52 $\pm$ 0.01 \\ \hline 
			$R_{q}$ [fm] 		& 0.26 $\pm$ 0.01 	&  0.25 $\pm$ 0.01 	&  0.25 $\pm$ 0.01 	&  0.28 $\pm$ 0.01 	&  0.39 $\pm$ 0.01 \\ \hline 
			$R_{d}$ [fm] 		& 0.71 $\pm$ 0.01 	&  0.71 $\pm$ 0.01 	&  0.71 $\pm$ 0.01 	&  0.75 $\pm$ 0.01 	&  0.83 $\pm$ 0.01 \\ \hline 
			$\chi^2/NDF$		& 65.0/44 			& 26.7/32 			& 57.2/33 			& 52.9/33 			& 4622.7/76  \\ \hline 
			$CL$ [\%] 			& 2.14 				& 73.05				& 0.55				& 1.56				& 0.0  \\ \hline 
			$\sigma_{elastic}$ [mb] & 6.2  $\pm$ 0.1  & 
				5.1  $\pm$ 0.3  & 
				5.0  $\pm$ 0.3  & 
				7.3  $\pm$ 0.3  & 
				17.9 $\pm$ 0.2 \\ \hline 
       	\end{tabular}
		\caption{The overall fit quality and resulting parameters of the fit at ISR energies including 
 	the LHC result at $7$ TeV. The diquark is assumed to be a single entity.}
		\label{parameters single}
	\end{table}

\vspace{1 truecm}
\vfill

\subsection{Model $p = (q, (q,q))$: diquark scatters as composite object}

\begin{figure}[H]
    	\includegraphics[width=0.9\textwidth]{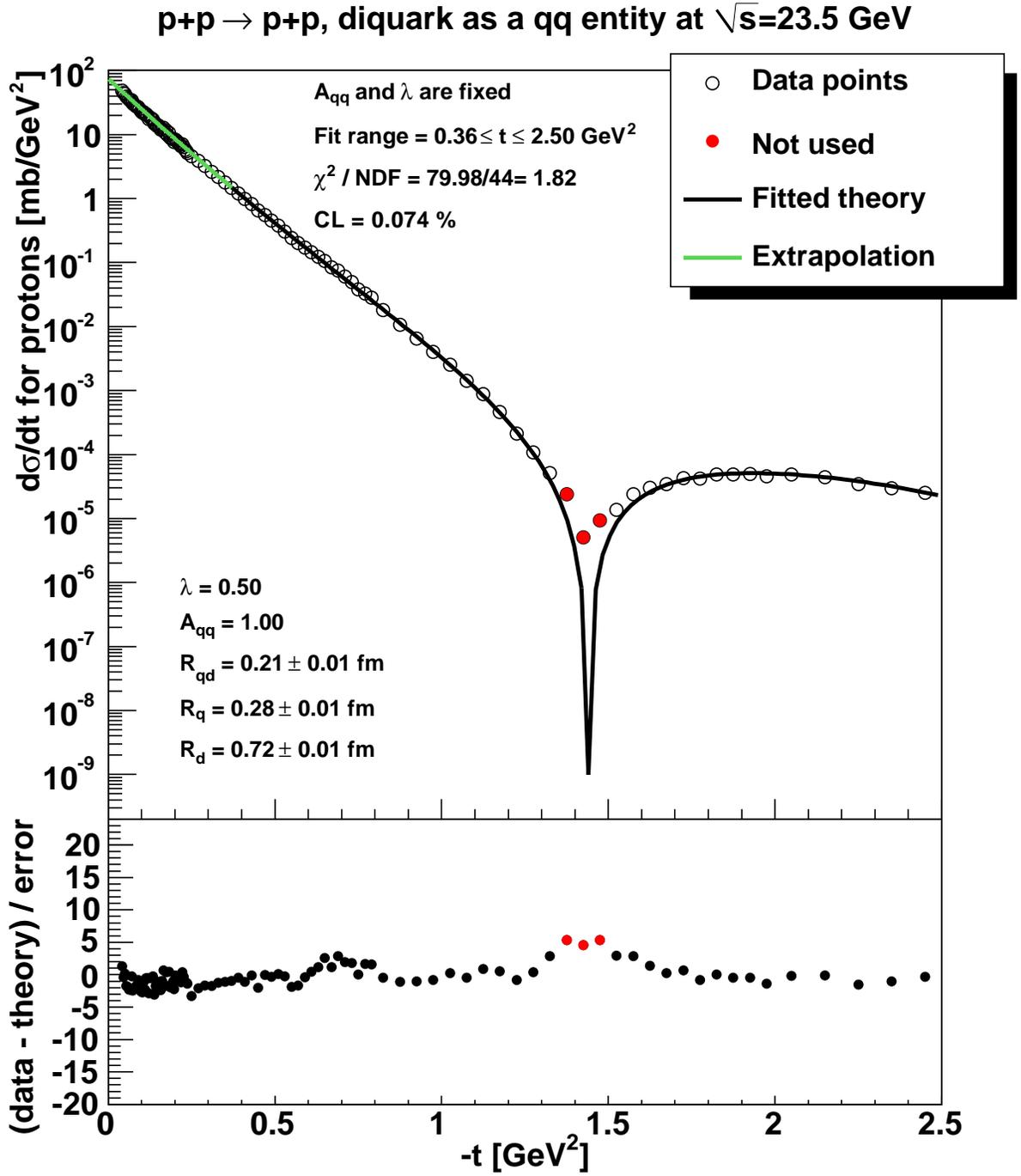}
        \centering
        \caption{
			(Color online.) Same as Fig. \ref{singlefitfor23}, but the diquark is assumed to have a composite $(q,q)$ substructure.
                }
		\label{qqfitfor23}
\end{figure}

\begin{figure}[H]
    	\includegraphics[width=0.9\textwidth]{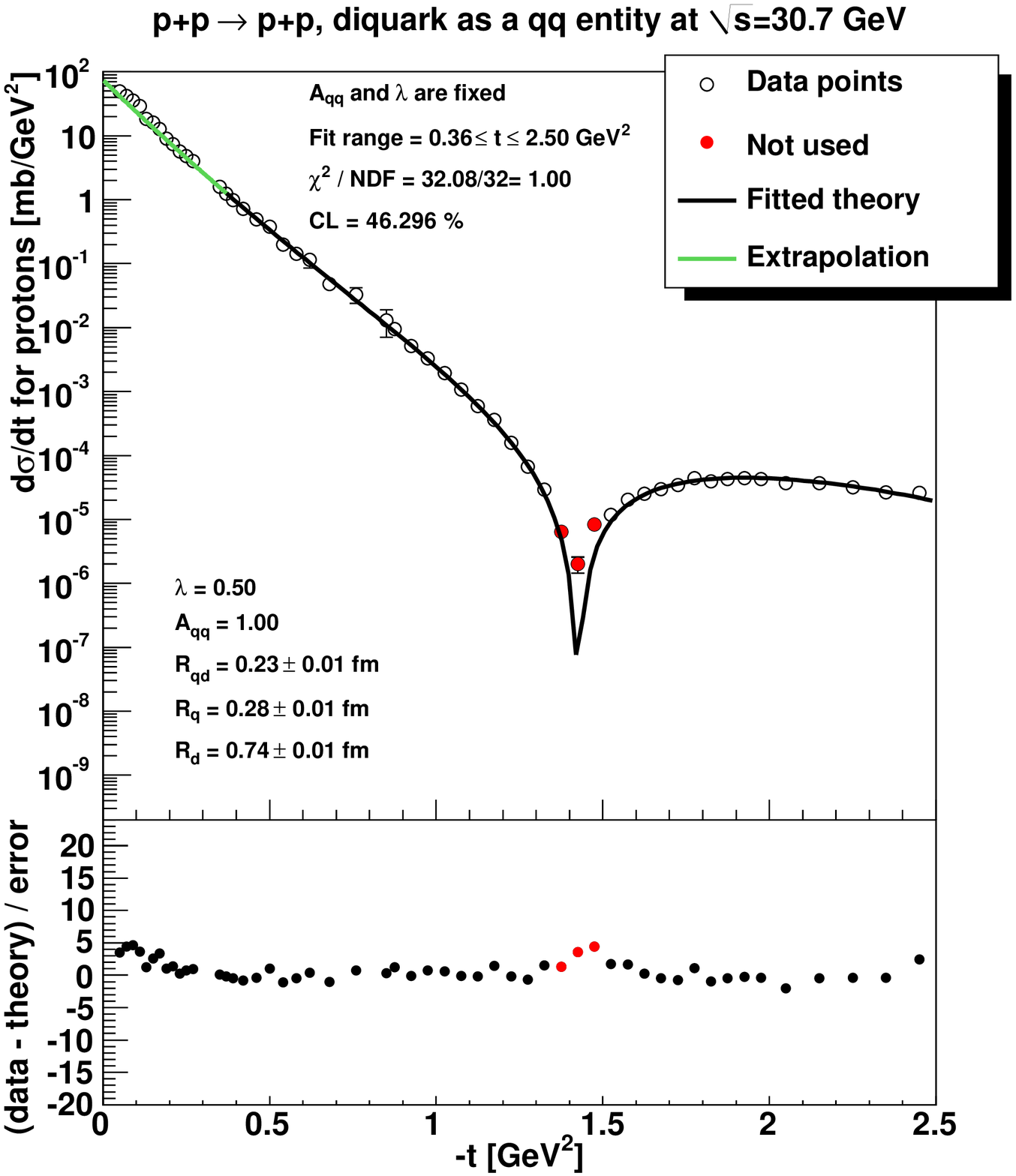}
        \centering
        \caption{(Color online.) Same as Fig \ref{qqfitfor23}, but for $\sqrt{s}=$ 30.7 GeV.}
		\label{qqfitfor31}
	\end{figure}

\begin{figure}[H]
    	\includegraphics[width=0.9\textwidth]{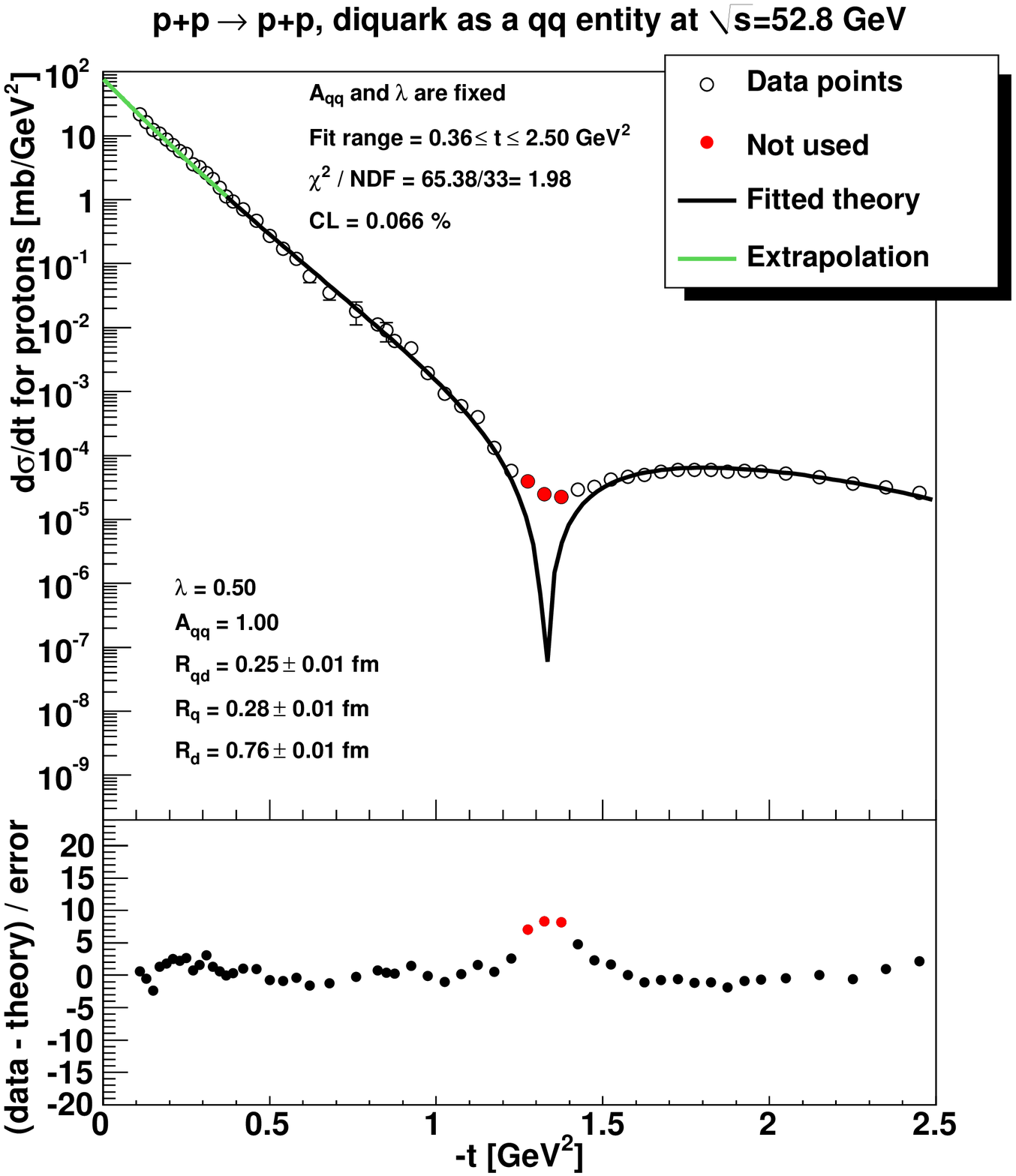}
        \centering
        \caption{(Color online.) Same as Fig \ref{qqfitfor23}, but for $\sqrt{s}=$ 52.8 GeV.}
		\label{qqfitfor53}
	\end{figure}

\begin{figure}[H]
		\includegraphics[width=0.9\textwidth]{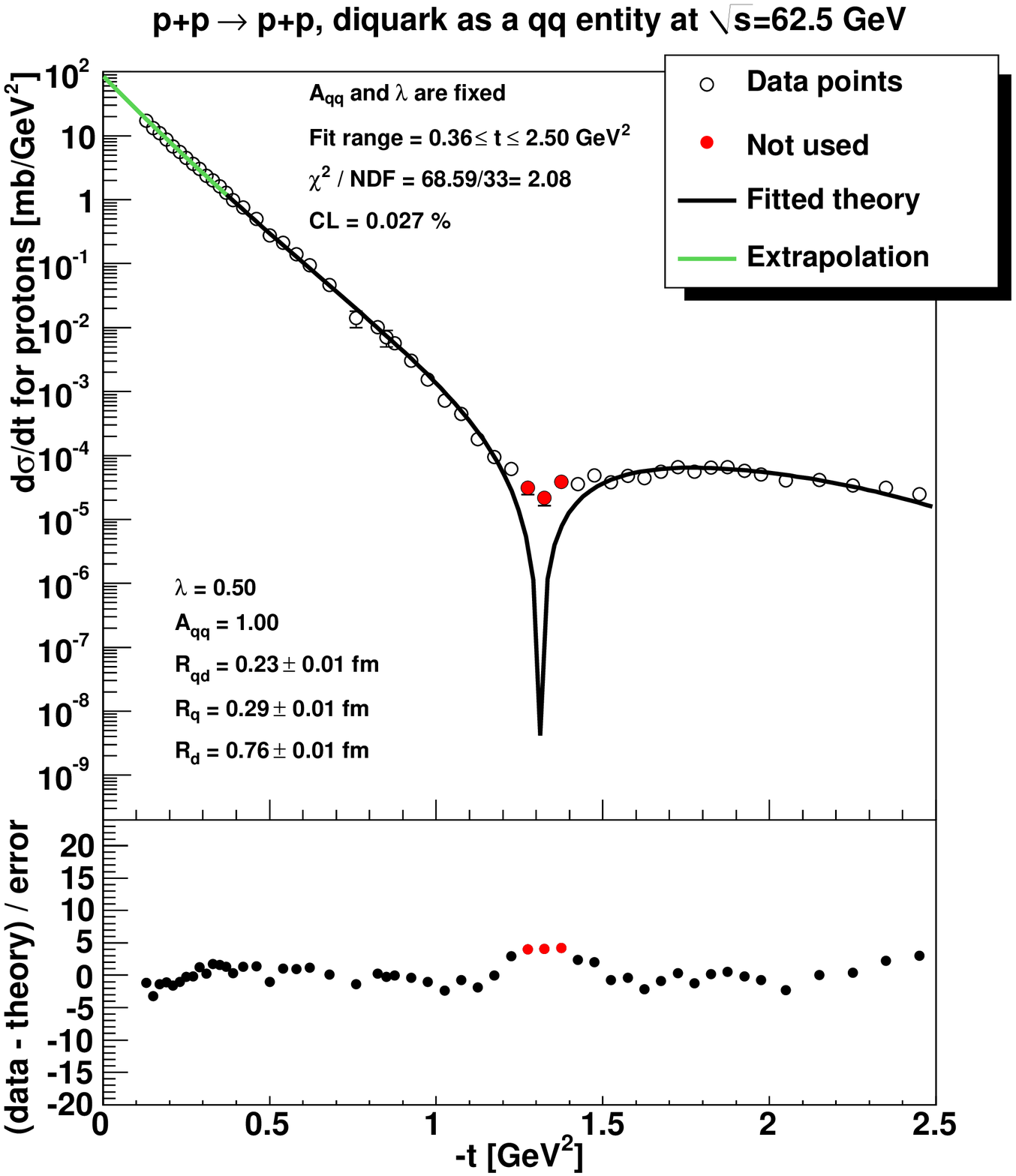}
		\centering
		\caption{(Color online.) Same as Fig \ref{qqfitfor23}, but for $\sqrt{s}=$ 62.5 GeV. 
        }
		
		\label{qqfitfor62}
        \end{figure}

\begin{figure}[H]
		\includegraphics[width=0.9\textwidth]{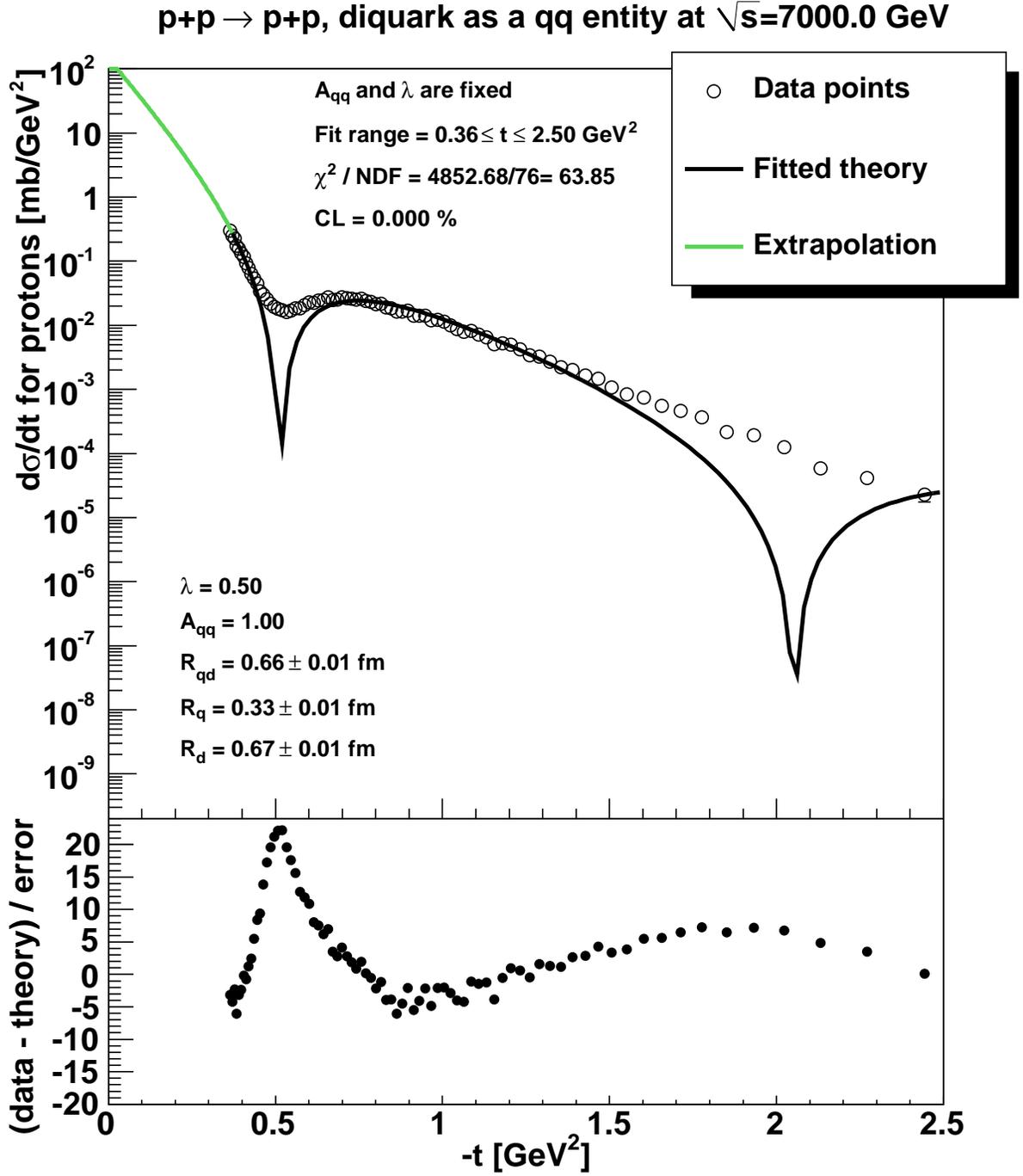}
		\centering
		\caption{
                (Color online.) Same as Fig \ref{qqfitfor23}, but for $\sqrt{s}=$ 7 TeV. 
                Note that this fit is not acceptable, since its CL is below 0.1\%.
		}
		\label{qq_fittotem}
	\end{figure}

In this subsection,  similar MINUIT fit results are presented as in the previous subsection,
the main modification is a change in the model assumption: now we assume that the diquark
scatters as a composite object that contains two quarks.
We present the final fit results to proton-proton elastic scattering data 
both at ISR \cite{Nagy:1978iw,Amaldi:1979kd}
and at LHC energies \cite{Antchev:2011zz}.
The results are illustrated on Fig. \ref{qqfitfor23}-\ref{qq_fittotem}. 
The confidence levels, and model parameters with their errors are summarized in Table \ref{tableqq}.

\vfill \newpage

	\begin{table}\footnotesize
    	\begin{tabular}{|c|c|c|c|c|c|} \hline
			$\sqrt{s}$ [GeV]& 23.5 				& 30.6 				& 52.8				&  62.5 			& 7000 \\ \hline\hline 
			$\lambda$ 		& 0.50 				&  0.50 			&  0.50 			&  0.50 			&  0.50  \\ \hline 
			$A_{qq}$ 		& 1.00 				&  1.00 			&  1.00 		 	&  1.00 			&  1.00  \\ \hline 
			$R_{qd}$ [fm] 	& 0.21 $\pm$ 0.01 	&  0.23 $\pm$ 0.01 	&  0.25 $\pm$ 0.01 	&  0.23 $\pm$ 0.01 	&  0.66 $\pm$ 0.01 \\ \hline 
			$R_{q}$ [fm] 	& 0.28 $\pm$ 0.01 	&  0.28 $\pm$ 0.01 	&  0.28 $\pm$ 0.01 	&  0.29 $\pm$ 0.01 	&  0.33 $\pm$ 0.01 \\ \hline 
			$R_{d}$ [fm] 	& 0.72 $\pm$ 0.01 	&  0.74 $\pm$ 0.01 	&  0.76 $\pm$ 0.01 	&  0.76 $\pm$ 0.01 	&  0.67 $\pm$ 0.01 \\ \hline 
			$\chi^2/NDF$ 	& 80.0/44 			& 32.1/32 			& 65.4/33 			& 68.6/33 			& 4852.7/76 \\ \hline 
			$CL$ [\%] 		& 0.07				& 46.30 			& 0.07				& 0.03 				& 0.0  \\ \hline 
			$\sigma_{elastic}$ [mb] & 
			6.9  $\pm$ 0.1  & 
			6.6  $\pm$ 0.1  & 
			6.6  $\pm$ 0.1  & 
			7.2  $\pm$ 0.1  & 
			9.8  $\pm$ 0.1 \\ \hline 
		\end{tabular}
		\caption{The overall fit quality and resulting parameters of the fit at the ISR energies including the LHC result at $7$ TeV. The 
				diquark is  assumed to be a $d=(q,q)$ composit entity.}
		\label{tableqq}
	\end{table}


\subsection{Total cross sections in the $p=(q,(q,q))$ model: the diquark is considered to be a composite object. }

The total inelastic cross sections for the quark-quark, quark-diquark and diquark-diquark
subcollisions were analyzed on the basis of formula (\ref{inelastic cross sections}). The detailed results are
collected in Table \ref{table:tableinelasticsigma}, while the average of these ratios for the described ISR energies are

	\begin{center}
    \begin{table}\footnotesize
        \begin{tabular}{|c|c|c|c|c|c|} \hline
            $\sqrt{s}$ [GeV]& 23.5 & 30.6 & 52.9 &  62.5 & 7000 \\ \hline\hline
            $\sigma_{qd}/\sigma_{qq}$ & 1.92 $\pm$ 0.01 &  1.93 $\pm$ 0.01 & 1.93  $\pm$ 0.01 & 1.93 $\pm$ 0.01 &  1.88 $\pm$ 0.01 \\ \hline
            $\sigma_{dd}/\sigma_{qq}$ & 3.64 $\pm$ 0.02 &  3.66 $\pm$ 0.01 & 3.67  $\pm$ 0.01 & 3.65 $\pm$ 0.01 &  3.43 $\pm$ 0.02 \\ \hline
        \end{tabular}
        \caption{The ratios of the total inelastic cross sections for the quark-quark, quark-diquark and 
		diquark-diquark processes for the ISR and 
 		LHC energies using the composite diquark hypothesis.}
        \label{table:tableinelasticsigma}
    \end{table}
	\end{center}

	\begin{center}
	\begin{table}\footnotesize
        \begin{tabular}{|c|c|c|c|c|c|} \hline
           $\sqrt{s}\,[GeV]$      	& 23.5 				& 30.6 				& 52.9				& 62.5	 			& 7000 \\ \hline 
           data   [mb]				& 38.94 $\pm$ 0.17 	& 40.14 $\pm$ 0.17 	& 42.67 $\pm$ 0.19 	& 43.32 $\pm$ 0.23 	& 98.3 $\pm$ 0.2 	\\ \hline \hline
		\multicolumn{6}{|c|}{p=(q,d)} \\\hline
           $\sigma_{total}$ [mb]		& 38.5  $\pm$ 0.2  	& 40.0  $\pm$ 0.2  	& 42.5  $\pm$ 0.2  	& 43.2  $\pm$ 0.3  	& 79.0 $\pm$ 1.1	\\ \hline
           $\chi^2$/NDF				& 107.2/45			& 62.2/33			& 110.7/34			& 56.1/34			& 4667.67/77 		\\ \hline
           CL     [\%]				& 0.00				& 0.16				& 0.00				& 1.00				& 0.00				\\ \hline
		\multicolumn{6}{|c|}{p=(q,(q,q))} \\\hline
           $\sigma_{total}$ [mb]		& 38.6  $\pm$ 0.1 	& 39.7  $\pm$ 0.2  	& 41.6  $\pm$ 0.2  	& 42.6  $\pm$ 0.2  	& 74.2 $\pm$ 1.3	\\ \hline 
           $\chi^2$/NDF				& 86.8/45			& 60.2/33			& 154.2/34 			& 95.3/34			& 5059.57/77			\\ \hline
           CL     [\%]				& 0.02				& 0.26				& 0.00				& 0.00				& 0.00				\\ \hline
        \end{tabular}
        \caption{Measured and fitted values of the  total cross sections,
        when the measured values of the total cross sections  are {\it included} into 
		the earlier fitting procedure,  as another, additional data point.}
        \label{table:results_sigmatotal_fit}
	\end{table}
	\end{center}

\begin{equation}
\sigma_{qq}:\sigma_{qd}:\sigma_{dd} = 1 :  (1.93 \pm 0.03) : (3.65 \pm 0.1), 
\label{ISRratios}
\end{equation}
which is close to the ideal 1: 2 : 4 ratio, confirming the assumption of having two quarks inside 
the diquark, amended with some shadowing which is 4\% and 9\% respectively. 
At 7 TeV the ratios are different from (\ref{ISRratios})  
\begin{equation}
	1 : (1.88 \pm 0.01) : (3.43 \pm 0.02),
	\label{ratiosat7TeV}
\end{equation}
which shows that shadowing is stronger, 6\% and 14\% percent respectively, and a significant decrease
compared to the ideal ratio can be observed.\par

The total elastic scattering cross sections were also determined, as given in Table \ref{tableqq}. 
In a preliminary conference proceedings~\cite{Nemes:2012mq}, we have noted that
that the errors on the total elastic cross sections might
be decreased by fixing the $A_{qq}$ and $\lambda$ parameters to their nominal values of 1 and 0.5, respectively.
Our final results presented here confirm this conjecture.
\vfill
\eject

		\begin{figure}[H]
		\includegraphics[trim = 6mm 12mm 2mm 4mm, clip, width=0.45\textwidth]{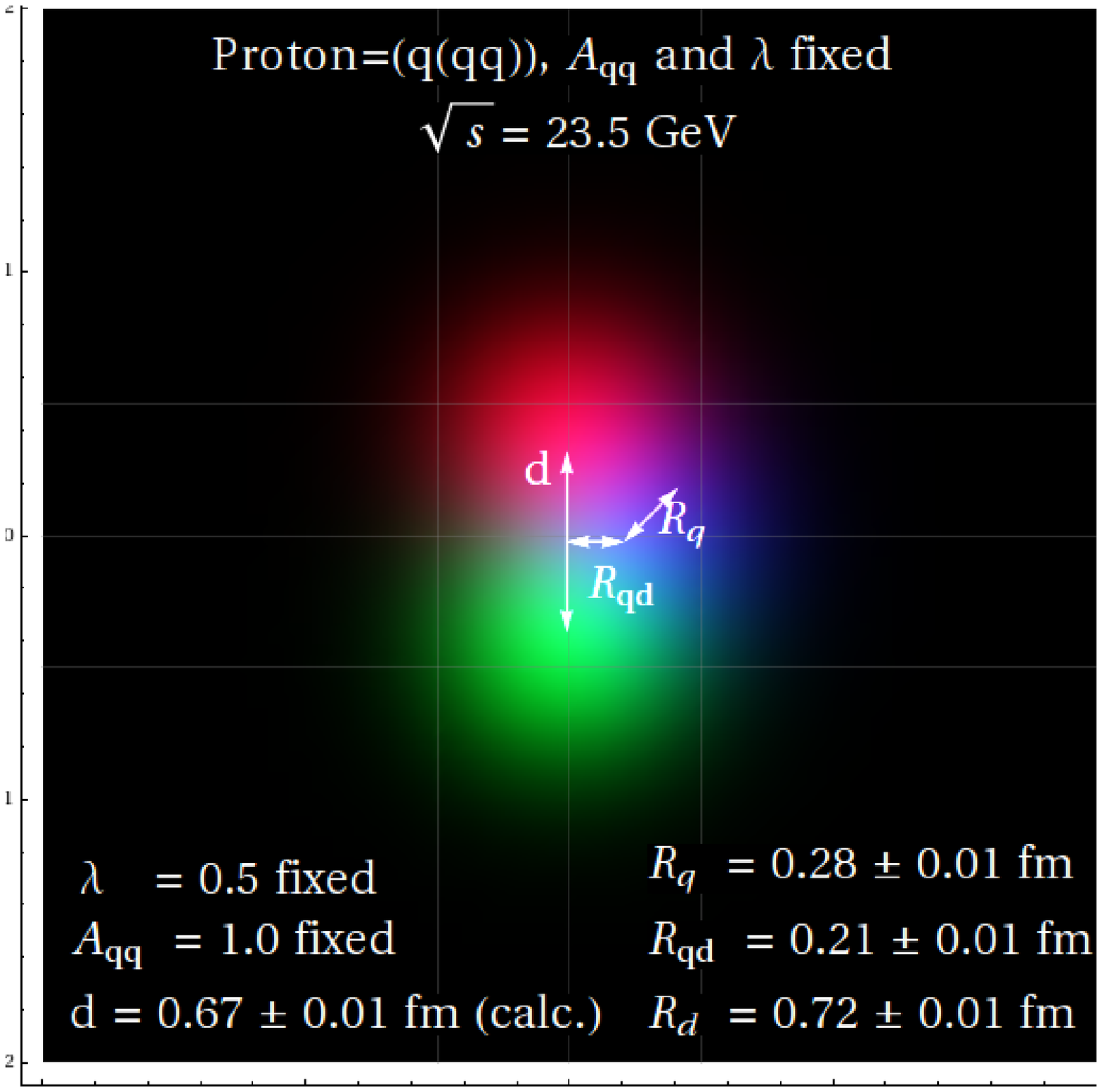}
    	   	\includegraphics[trim = 6mm 12mm 2mm 4mm, clip, width=0.45\textwidth]{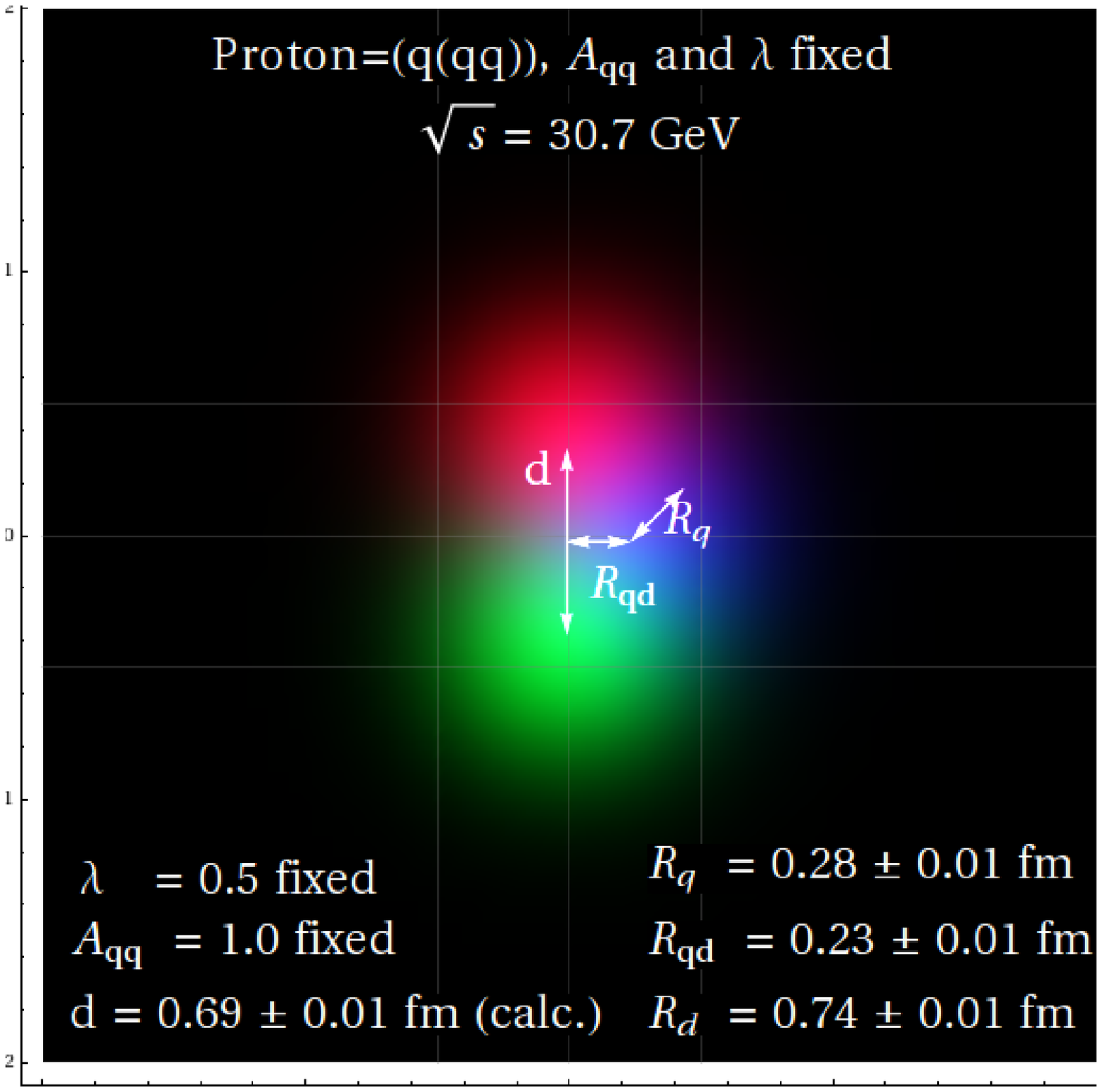}
	        \includegraphics[trim = 6mm 12mm 2mm 4mm, clip, width=0.45\textwidth]{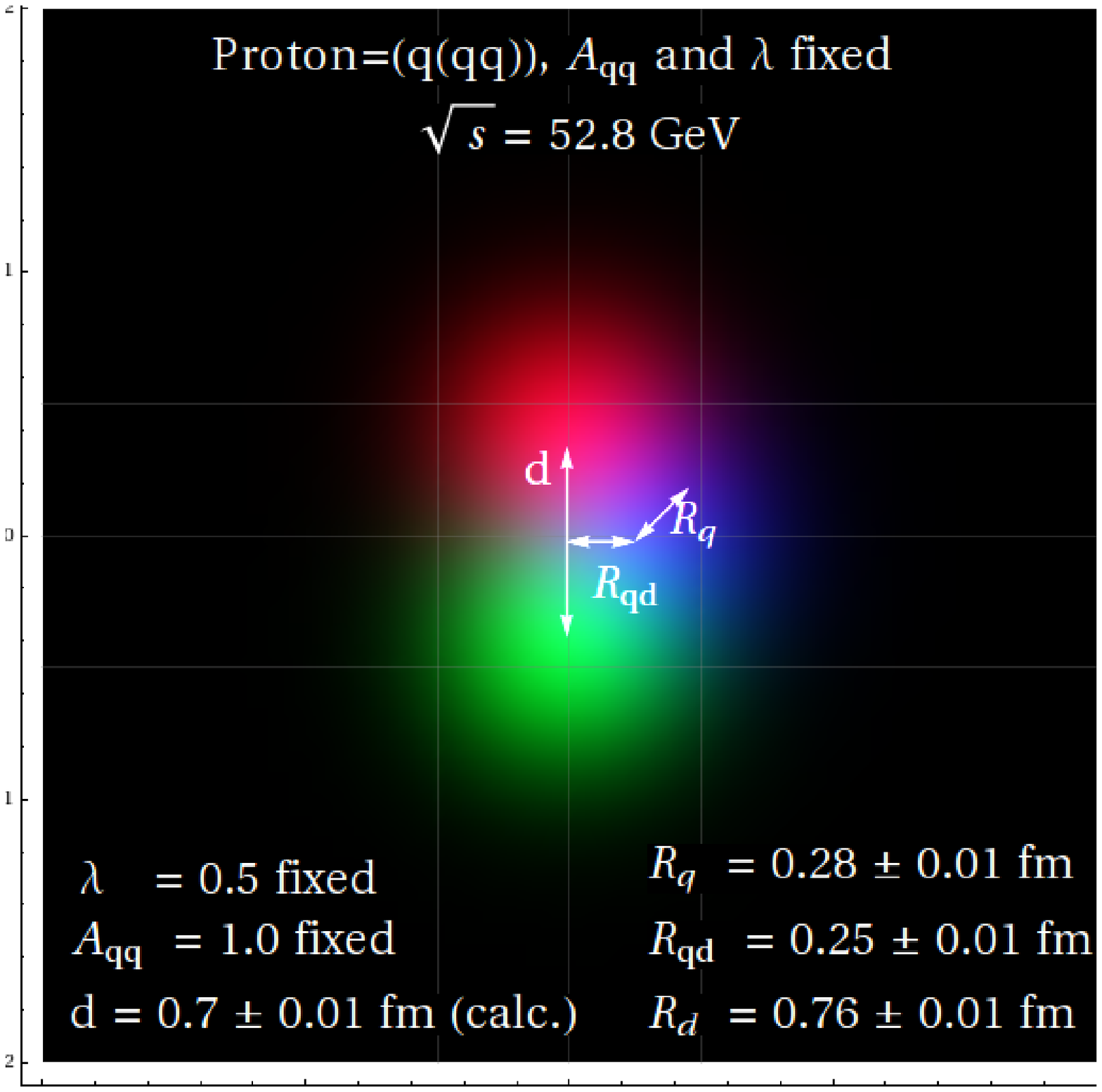}
	    	\includegraphics[trim = 6mm 12mm 2mm 4mm, clip, width=0.45\textwidth]{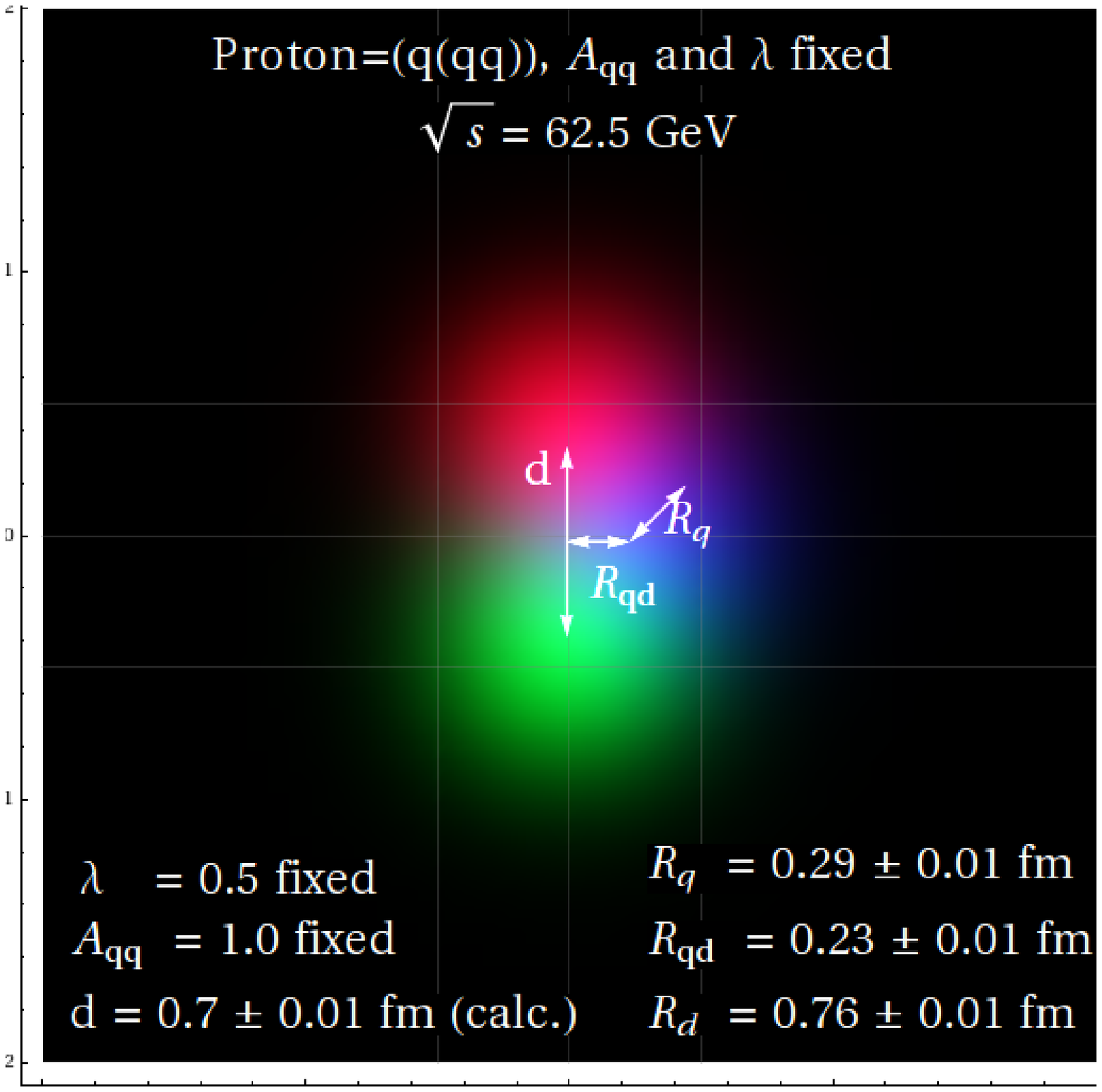}
    	    	\includegraphics[trim = 6mm 12mm 2mm 4mm, clip, width=0.45\textwidth]{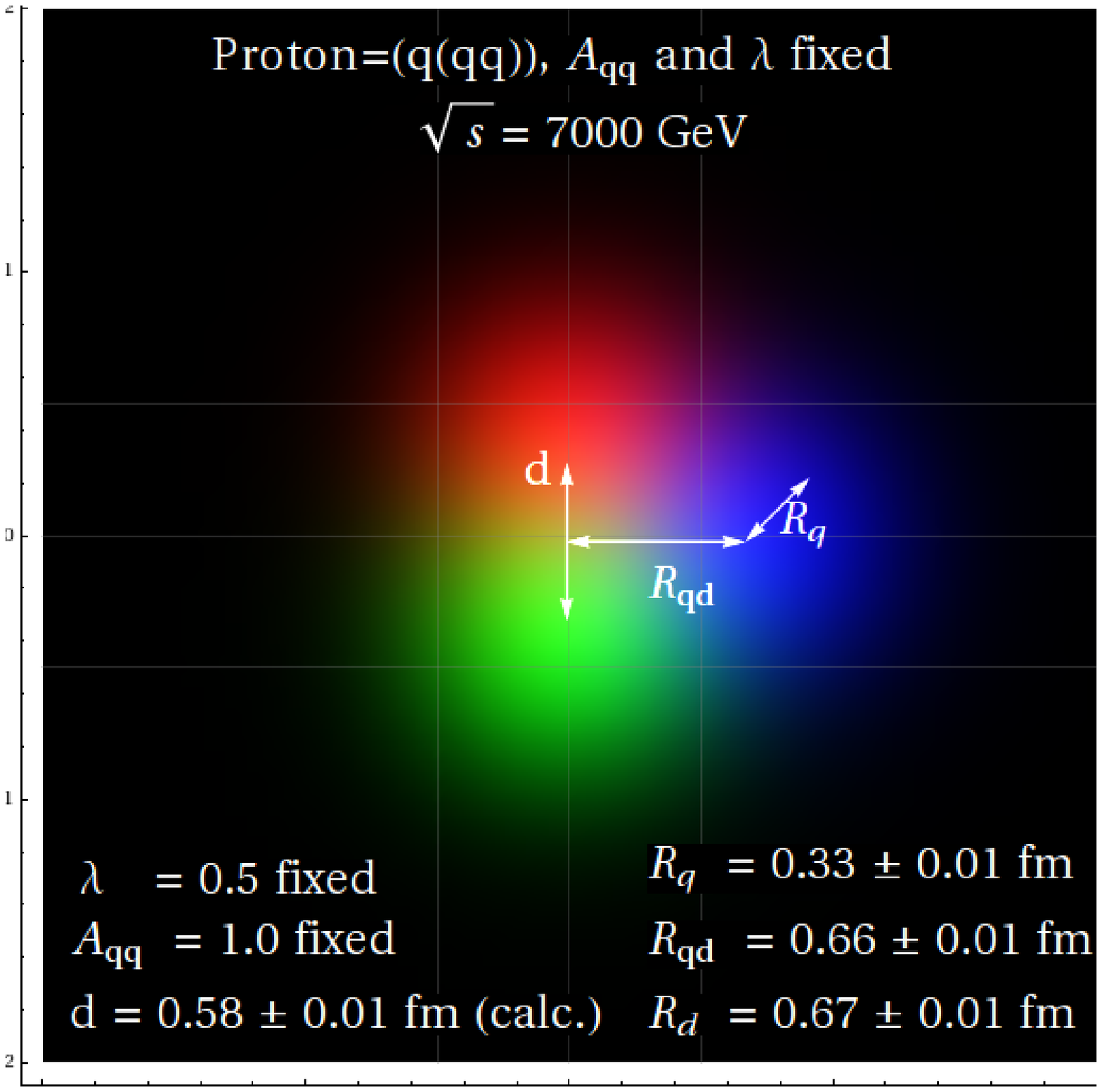}
			\centering
			\caption{(Color online.) Visualisation of the obtained $R_{qd}$, $R_q$, $R_d$ parameters when the diquark is assumed to be a {\it qq} entity.
				 The proton seems to be much larger at LHC energies than in the ISR regime. 
				This is mainly due to an increase in the $R_{qd}$ parameter,
				that characterizes the separation of the quark and the diquark, which is further decomposed as (q,q) state 
				in this class of models.
                 }
			\label{visualisationqq}
	    \end{figure}

\subsection{Model comparison}

\begin{figure}[H]
        \centering
        \includegraphics[width=0.90\textwidth]{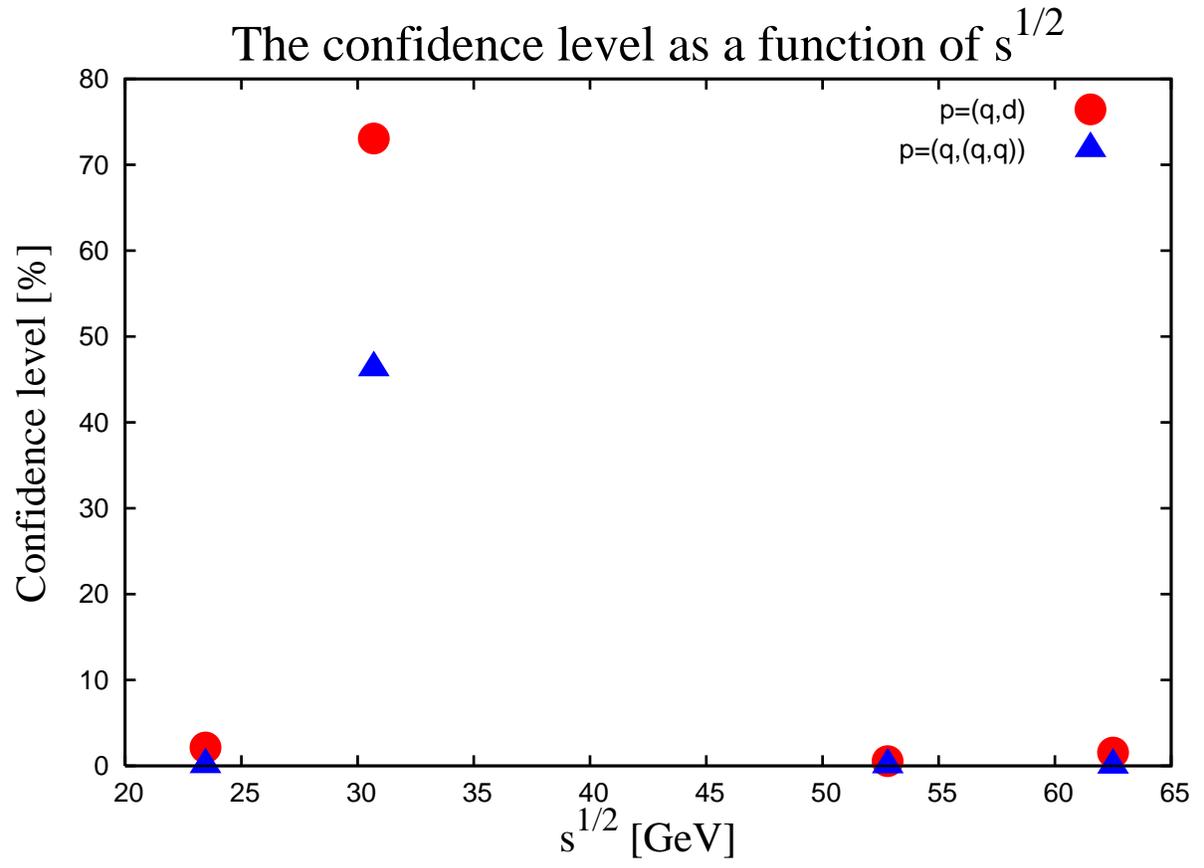}
        \caption{(Color online.) The confidence level as a function of $\sqrt{s}$.
        The value of CL at 7 TeV is practically zero and it is not shown. 
        }
        \label{confidencesingle}
    \end{figure}

    The $\sqrt{s}$ dependence of the parameters is summarized here. The confidence levels as a
    function of the center of mass energy for both of
    the models are given on Fig. \ref{confidencesingle}. We left out the confidence level for TOTEM
    which is zero, all the other values are higher than or close to $0.1\%$.\par
\begin{figure}[H]
		\centering
		\includegraphics[width=0.90\textwidth]{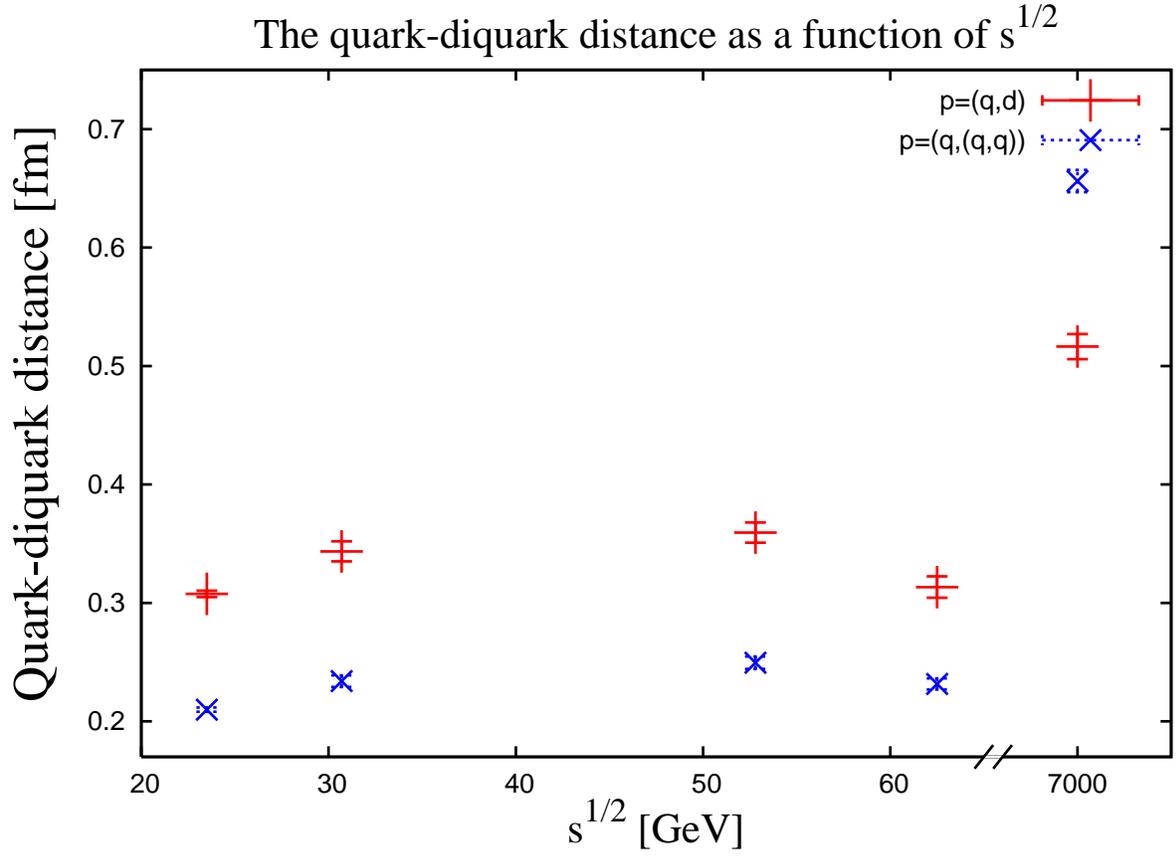}
		\caption{(Color online.) Parameter $R_{qd}$, representing the separation of the quark and the diquark, 
        as a function of $\sqrt{s}$.}
		\label{Rp}
	\end{figure}
\begin{figure}[H]
        \includegraphics[width=0.90\textwidth]{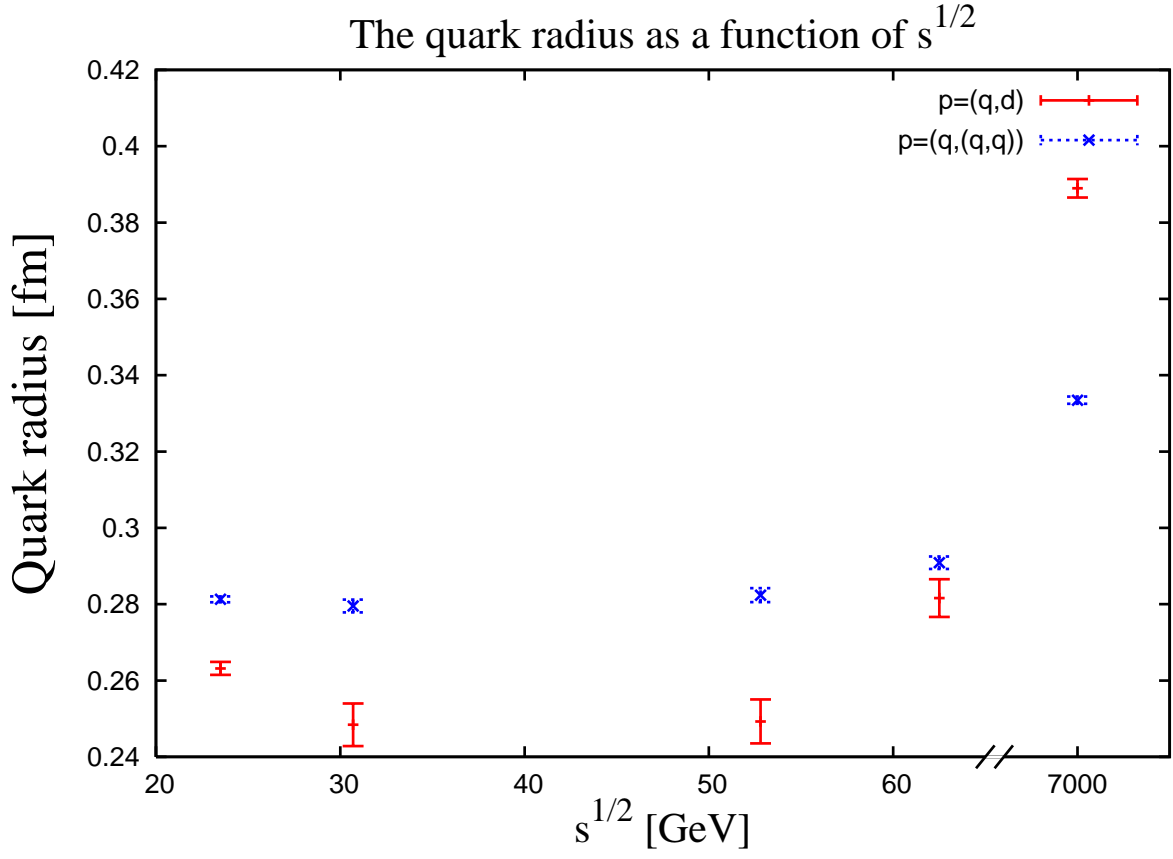}
		\caption{(Color online.) Parameter $R_q$ (quark size) as a function of $\sqrt{s}$.}
		\label{Rq}
    \end{figure}
\begin{figure}[H]
		\centering
                \includegraphics[width=0.90\textwidth]{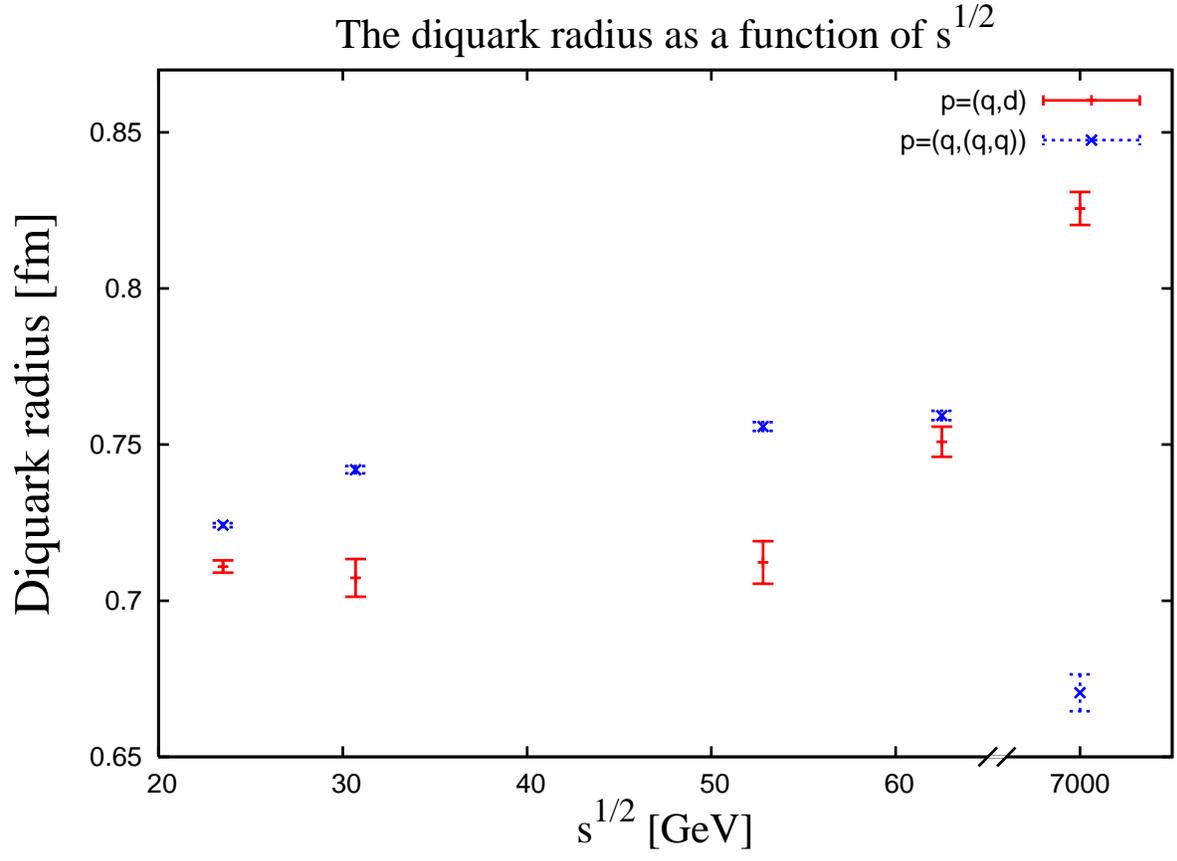}
		\caption{(Color online.) Parameter $R_d$, the diquark size, as a function of $\sqrt{s}$.}
		\label{Rd}
    \end{figure}

    The center of mass energy dependence of the quark-diquark distance is presented on Fig. \ref{Rp}. 
        The effective quark radius can be seen on Fig. \ref{Rq} and the obtained diquark radius is on
        Fig. \ref{Rd}.

	As can be seen on these figures, the fit quality is similar and the best fit parameters are rather different for both models at each colliding energies and the picture of the protons as realized by Figs.
    \ref{single visualisation} and \ref{visualisationqq} are rather different. We started to wonder how it is possible, that the apparently rather different p=(q,d) and p=(q,(q,q)) models give so similar fit results, 
	not only quantitatively but qualitatively too? Although the mathematical equations that describe the $d\sigma/dt$ from the two models are 
	rather complex and formally different from one another, we investigated some simple relationships among the model parameters. 
	\par
\begin{figure}[H]
        \centering
                \includegraphics[width=0.90\textwidth]{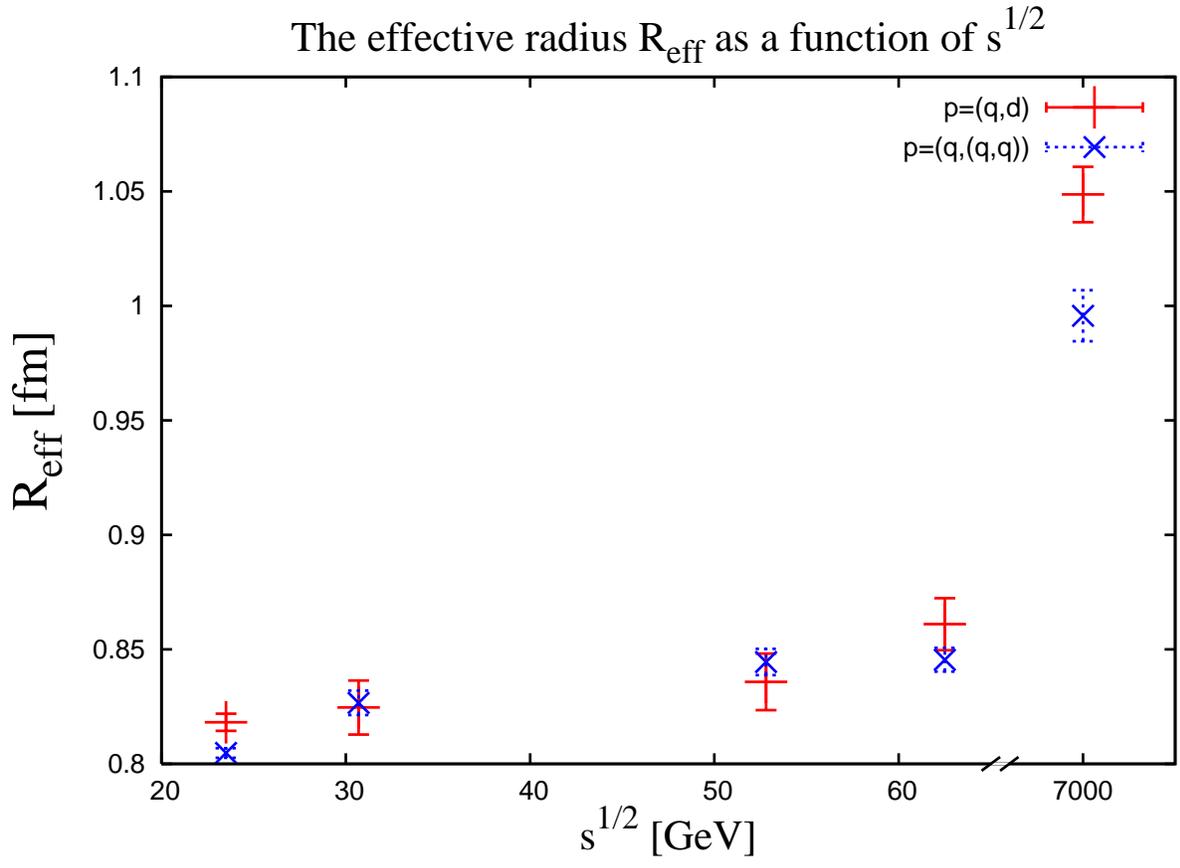}
        \caption{(Color online.) The model independent $R_{\mbox{\rm\scriptsize eff}}$ radius parameter as a function of $\sqrt{s}$.}
        \label{Reff}
    \end{figure}

\begin{figure}[H]
        \centering
                \includegraphics[width=0.90\textwidth]{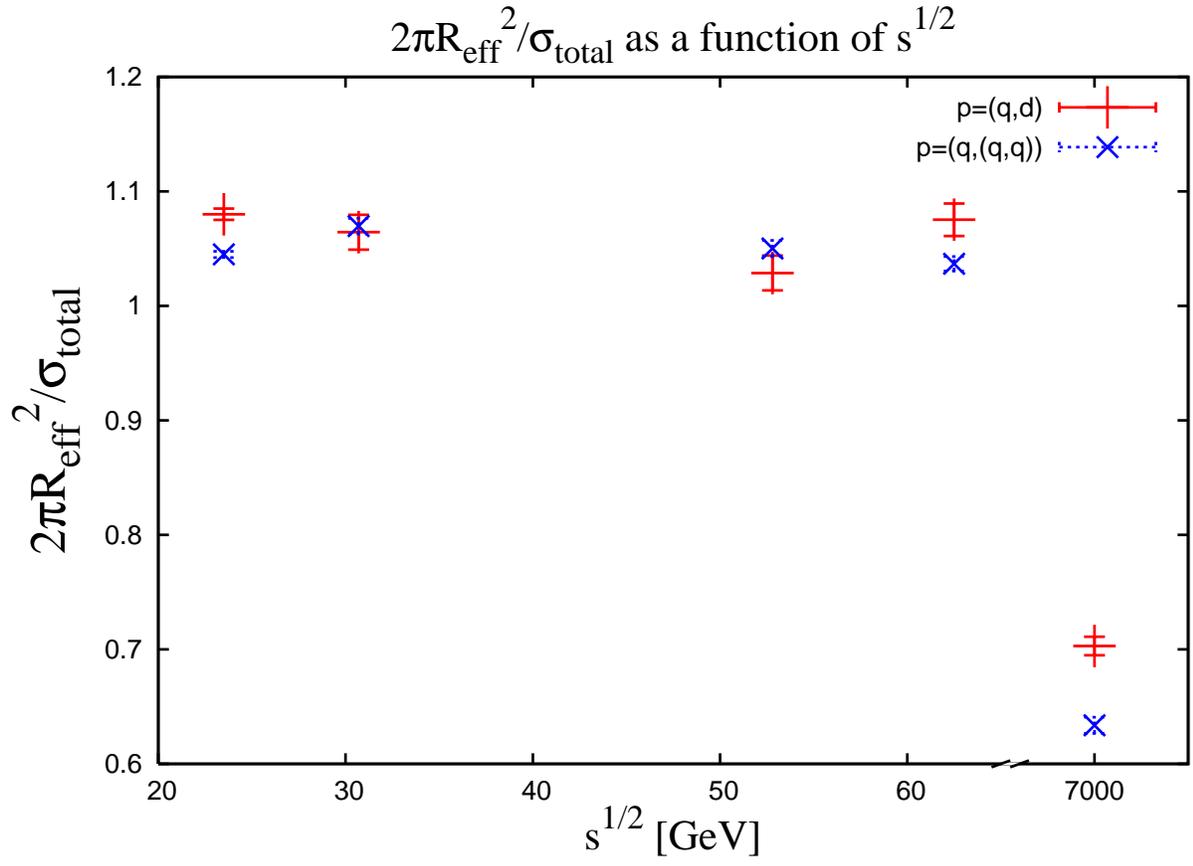}
        \caption{(Color online.) The effective $R_{\mbox{\rm\scriptsize eff}}$ radius divided by the measured total cross section as a function of $\sqrt{s}$. The
		result indicates a simple relationship between these parameters, namely, both for the case when a proton is modelled as p = (q,d) and 
		for the case when p = (q, (q,q)), the quadratic sum
		of the characteristic radii is the same model independent quantity.}
        \label{Reffpersigmatot}
    \end{figure}

 
	Numerically, we found one combination of the model parameters, 
	an effective radius that is
	obtained from the quadratic sum of $R_q$, $R_d$ and $R_{qd}$, that seems to be the same in both models and which is related in a simple way to 
	the measured total cross-section.  
	\begin{align}
		R_{\mbox{\scriptsize \rm eff}} = \sqrt{R_{q}^2 + R_{d}^2 + R_{qd}^2}\;,
		\label{eqReff}
	\end{align}
	\begin{align}
		\sigma_{total}=2 \pi R_{\mbox {\rm\scriptsize eff}}^2\;.
		\label{eqReffpersigmatot}
	\end{align}
	The $\sqrt{s}$ dependence of the effective radius $R_{\mbox{\scriptsize \rm eff}}$ is shown on Figure \ref{Reff}, and
	the relationship between the effective radius and the measured total cross section is presented on 
	Figure \ref{Reffpersigmatot}. 
\vfill
\eject

\newpage
\section{Conclusion and outlook}
\label{sec:conclusion}
	A systematic study of fit quality as well as the fit parameters under similar 
    circumstances have been performed for the Bialas - Bzdak model~\cite{Bialas:2006qf}
    in a wide energy
 	range from ISR to LHC energies using the same kinematic interval and the same method at each energy. The model gives a good description of the 
	ISR data, which means that the CL is acceptable on the
	ISR energies if 3 data points at first diffractive minimum were left out from the fit. 
    These studies confirm earlier results \cite{Bialas:2006qf} concerning the increase of the  total proton-proton cross section 
    (``size'' of the proton) with increased colliding energies.

    An important shortcoming of the quark-diquark model of protons is that it ignores the real part
    of the elastic scattering amplitude. This leads to a singular behaviour at the diffractive minimum,  which is apparently a more and more serious model 
	limitation with increasing the energy of p+p collisions.
    Due to this reason, the considered model fails to describe in detail
    the structure of the first diffractive minimum at the LHC energies of $7$ TeV. 

	We found a combination of the model parameters, an effective radius that is obtained as the quadratic sum of quark, diquark radii and 
	the separation between the quark and the center of mass of the diquark, that seems to be the same in both models and which is
	related in a simple, intuitive way to the measured total cross-section, as given by eqs. (\ref{eqReff}) and (\ref{eqReffpersigmatot}). As
 	 indicated on Figure \ref{Reffpersigmatot}, the precision of this "rule of thumb" formula 
	is about 10 \% which is quite amazing for us given that
	it is an extremely simple formula as compared to the 
	full, exact and analytic expressions that are also obtained from both models.

    The evaluated ratios of the quark-quark, quark-diquark and diquark-diquark total inelastic cross-sections were found to deviate more and more from 
	the ideal 1 : 2 : 4 ratio with increasing energies. In the ISR energy range the deviations from this ideal value were less than a $5 \sigma$ effect, 
indicating lack of significant shadowing effects. However at the current LHC energy of $\sqrt{s} = 7$ TeV, a significant decrease -- as 
	indicated by eq. (\ref{ratiosat7TeV}) -- compared to these ideal ratios were found, which possibly may indicate an increased role of shadowing
    at CERN LHC energies.

    Finally let us note that the TOTEM Collaboration extended recently the measurement of
    the differential elastic p+p scattering cross-sections to low values of $|t|$
    in ref.~\cite{Antchev:2011vs}, allowing one
    to extrapolate to the optical point at $t=0$ and to determine the total elastic and the total
    scattering cross-sections of p+p collisions at $\sqrt{s} = 7$ TeV for the first time, but
    these data were yet not utilized in our analysis.

\section{Acknowledgements}
{ T. Cs. would like to thank prof. R. J. Glauber for valuable discussions at the initial phase of this project, and for his kind hospitality at Harvard University
and to I. Dremin for inspiring discussions.
F. N. would like to thank A. Ster and M. Csan\'ad for their valuable help with the CERN MINUIT multi-parameter optimization. The authors would like to thank R. 
Lohner for a careful reading of the draft of this manuscript. This research was partially supported
by the Hungarian American Enterprise Scholarship Fund (HAESF) and by the Hungarian OTKA grant NK 101438.
}

\end{document}